\documentclass[aps,pre,reprint,twocolumn,superscriptaddress]{revtex4-1}

\pdfoutput=1

\usepackage{graphicx}
\usepackage[caption=false]{subfig}
\usepackage{amsmath}
\usepackage{amssymb}
\usepackage{color}

\DeclareMathOperator{\Tr}{Tr}	

\begin{document}


\title{Displacement field around a rigid sphere in a compressible elastic environment, corresponding higher-order Fax\'en relations, as well as higher-order displaceability and rotateability matrices}

\author{Mate Puljiz}
\email{puljiz@thphy.uni-duesseldorf.de}
\affiliation{Institut f{\"u}r Theoretische Physik II: Weiche Materie, 
Heinrich-Heine-Universit{\"a}t D{\"u}sseldorf, D-40225 D{\"u}sseldorf, Germany}
\author{Andreas M. Menzel}
\email{menzel@thphy.uni-duesseldorf.de}
\affiliation{Institut f{\"u}r Theoretische Physik II: Weiche Materie, 
Heinrich-Heine-Universit{\"a}t D{\"u}sseldorf, D-40225 D{\"u}sseldorf, Germany}

\date{\today}

\begin{abstract}
An efficient route to the displacement field around a rigid spherical inclusion in an infinitely extended homogeneous elastic medium is presented in a slightly alternative way when compared to some common textbook methods. Moreover, two Fax\'en relations of next-higher order beyond the stresslet are calculated explicitly for compressible media. They quantify higher-order moments involving the force distribution on rigid particles in a deformed elastic medium.
Additionally, the displaceability and rotateability matrices are calculated up to (including) sixth order in inverse particle separation distance. These matrices describe the interactions mediated between the rigid embedded particles by the elastic environment. All methods and results can formally be transferred to the corresponding case of incompressible hydrodynamic low-Reynolds-number Stokes flow by considering the limit of an incompressible environment. The roles of compressibility of the embedding medium and of the here additionally derived higher-order contributions are highlighted by some selected example configurations.
\end{abstract}



\maketitle

\section{Introduction}

A basic analytical solution provided by low-Reynolds-number hydrodynamics concerns the flow past a uniformly translating or rotating rigid spherical particle in an incompressible, infinitely extended, and otherwise quiescent fluid.
It is explicitly treated by many classical textbooks, see, for example, Refs.~\onlinecite{happel1981low,karrila1991microhydrodynamics,dhont1996introduction}.
The flow field that is created around the sphere under no-slip surface conditions is referred to as Stokes flow.
A complementary problem concerns the situation, in which a sphere is immersed in a fluid, where it is subject to a given flow field. 
The sphere is then translated and rotated by the flow.
In this case, to find the translational and angular velocity of the particle, the so-called Fax\'en relations may be employed \cite{batchelor1972hydrodynamic,happel1981low,karrila1991microhydrodynamics,dhont1996introduction}.
These relations allow to determine the velocity, the angular velocity, and the stresslet that a rigid sphere acquires in an arbitrary imposed flow field.
The solutions to both problems are important tools to calculate the hydrodynamic interactions between the spheres within semidilute many-sphere systems in the form of mobility matrices \cite{karrila1991microhydrodynamics,dhont1996introduction,mazur1982many}.
Refs.~\onlinecite{happel1981low,karrila1991microhydrodynamics,dhont1996introduction} themselves cite many related works.

The formal similarity between the incompressible low-Reynolds-number hydrodynamics problem and the corresponding problem in linear elasticity theory has been noted several times.
In the hydrodynamic formulation \cite{happel1981low,karrila1991microhydrodynamics,dhont1996introduction}, there is only one material parameter, namely the shear viscosity $\eta$ of the incompressible fluid.
In linear elasticity theory, an infinitely extended and isotropic compressible homogeneous medium is described by two material parameters \cite{landau1986theory,mura1987micromechanics}.
First, the shear modulus $\mu$ is the elastic analogon to the hydrodynamic shear viscosity.
The second parameter determines the compressibility of the material, for instance, in the form of the Poisson ratio $\nu$.
It is thus possible to obtain the hydrodynamic expressions from the linearly elastic equations by assuming the elastic material to be incompressible, which formally corresponds to taking the limit of $\nu\rightarrow1/2$.
Then, passing to the hydrodynamic case, the elastic displacement field is replaced by the hydrodynamic flow field and the shear modulus $\mu$ by the hydrodynamic viscosity $\eta$.
The methods employed in the present work can in this way directly be transferred to hydrodynamic situations of incompressible fluid flows.

In the following, we address the two initially stated problems for compressible linear elasticity theory, thereby extending results in earlier works \cite{puljiz2016forces,puljiz2017forces}.
First, we present a straightforward, slightly alternative approach to the well-known displacement field created by a uniformly translated and rotated sphere, which we have not found in the mentioned textbooks of hydrodynamics (for the Stokes flow past a sphere) nor in corresponding literature for linear elasticity theory \cite{mura1987micromechanics,phanthien1993rigid}.
Here, the framework of the multipole expansion is used in combination with an ansatz for the force density
(compare, e.g., to Refs.~\onlinecite{karrila1991microhydrodynamics,phanthien1993rigid}, which use the multipole expansion in the same context, but rather present the final expressions that satisfy the boundary conditions and the underlying equations).
We use an ansatz for the surface force density on the sphere \cite{dhont1996introduction} and show how this ansatz, when inserted into the multipole expansion, leads to the familiar results that satisfy the requirements on the solutions.
As for the second problem, we calculate the third- and fourth-rank Fax\'en laws beyond the stresslet for a compressible medium, following the route outlined in Ref.~\onlinecite{batchelor1972hydrodynamic} for the incompressible hydrodynamic case.
We have not found these expressions in Refs.~\onlinecite{karrila1991microhydrodynamics,happel1981low,dhont1996introduction,kim1995faxen,phanthien1994loadtransfer,mura1987micromechanics}.
As an application of these additional Fax\'en relations, the displaceability and rotateability matrices up to sixth order in inverse interparticle separation distance are calculated for a compressible or incompressible elastic environment.
They quantify the interactions between embedded rigid spherical particles as mediated by the distorted environment, including three-body interactions.
In the incompressible situation, the hydrodynamic mobility matrices as presented in Ref.~\onlinecite{mazur1982many} are formally recovered.
The difference due to compressibility is demonstrated by exemplary induced interactions between paramagnetic spherical particles embedded in an elastic environment, which has already to lower order found application in the theoretical description of elastic magnetic gels \cite{puljiz2016forces,puljiz2018reversible,menzel2017force,schopphoven2018elastic,puljiz2019memory}.

In Sec.~\ref{oseen_multipole} the Green's function and its multipole expansion are presented.
The displacement field around a uniformly translated and rotated sphere is then calculated in Sec.~\ref{flow_past_sphere}.
In Sec.~\ref{faxen}, the two Fax\'en laws beyond the stresslet are derived.
The displacement field around a sphere that is both actively displaced and rotated by an external force and/or torque, respectively, as well as by a deformation of the surrounding medium is calculated.
Then, we determine the displaceability and rotateability matrices in a compressible medium up to sixth order in inverse interparticle distance in Sec.~\ref{matrixmediated} together with some simple examples that show the effect of compressibility and of the additionally derived higher-order contributions.
We conclude in Sec.~\ref{conclusion}.

\section{Green's function and multipole expansion}
\label{oseen_multipole}

In linear elasticity theory, the displacement field $\mathbf{u}(\mathbf{r})$ of an unbounded, homogeneous, and isotropic elastic medium is in general described by the Navier--Cauchy equation \cite{landau1986theory}
\begin{equation}\label{navier_cauchy}
	\nabla^2\mathbf{u}(\mathbf{r}) + \frac{1}{1-2\nu}\nabla\nabla\cdot\mathbf{u}(\mathbf{r}) + \frac{1}{\mu}\mathbf{f}_b(\mathbf{r}) ={} \mathbf{0},
\end{equation}
with $\nu$ denoting the Poisson ratio, $\mu$ the elastic shear modulus, and $\mathbf{f}_b(\mathbf{r})$ a force volume density.
From Eq.~(\ref{navier_cauchy}), the relations
\begin{eqnarray}
	\nabla^2\nabla\cdot\mathbf{u}(\mathbf{r}) &={} & 0,\label{laplace_div_u}	\\
	\nabla^4\mathbf{u}(\mathbf{r}) &={} & \mathbf{0},
	\label{biharmonic}	\\
	\nabla\times\nabla^2\mathbf{u}(\mathbf{r}) &={} & \mathbf{0}
	\label{rot_condition}
\end{eqnarray}
can be deduced for areas of constant or vanishing $\mathbf{f}_b(\mathbf{r})$ \cite{puljiz2017forces}.
Equation (\ref{biharmonic}) is also referred to as the biharmonic equation.

The displacement field induced by a constant point force $\mathbf{F}$ attacking at the origin, i.e., $\mathbf{f}_b(\mathbf{r})=\mathbf{F}\delta(\mathbf{r})$, can be written in the form
$\mathbf{u}(\mathbf{r})=\mathbf{\underline{G}}(\mathbf{r})\cdot\mathbf{F}$.
The corresponding Green's function $\mathbf{\underline{G}}(\mathbf{r})$ is a symmetric second-rank tensor and reads \cite{landau1986theory,puljiz2017forces}
\begin{equation}{\label{greens_function}}
	\mathbf{\underline{G}}(\mathbf{r})={}\frac{1}{16\pi(1-\nu)\mu r}\left((3-4\nu) \mathbf{\underline{\hat{I}}} + \mathbf{\hat{r}}\mathbf{\hat{r}} \right),
\end{equation}
with $\mathbf{\underline{\hat{I}}}$ denoting the identity matrix, $r=|\mathbf{r}|$, $\mathbf{\hat{r}}=\mathbf{r}/r$, and $\mathbf{\hat{r}}\mathbf{\hat{r}}$ representing a dyadic tensor.
This solution satisfies Eqs.~(\ref{navier_cauchy})--(\ref{rot_condition}).

In the case of a spatially extended, displaced rigid particle in an otherwise unperturbed elastic medium, the displacement field can be written as
\begin{equation}\label{Multipole-foundation}
	\mathbf{u}(\mathbf{r}) ={} \int_{\partial V} \mathrm{d}S' ~ \mathbf{\underline{\hspace{-.02cm}G}}(\mathbf{r}-\mathbf{r}')\cdot\mathbf{f}(\mathbf{r}'),
\end{equation}
where $\mathbf{r}'$ are the points on the particle surface $\partial V$ and $\mathbf{f}(\mathbf{r}')$ is the force surface density exerted by the particle onto the medium \cite{phanthien1993rigid,puljiz2017forces}.
Equation (\ref{Multipole-foundation}) follows from Eq.~(\ref{greens_function}) via the superposition principle due to the linearity of the Navier--Cauchy equation.
Thus, the displacement field is solely induced by the translating and/or rotating particle shell acting on the medium.

Switching to index notation and considering a particle located around the origin at $\mathbf{r}=\mathbf{0}$, the Taylor series of $\mathbf{\underline{\hspace{-.02cm}G}}(\mathbf{r}-\mathbf{r}')$ in $\mathbf{r}'$ reads 
\begin{equation}
	G_{ij}(\mathbf{r}-\mathbf{r}') ={} \sum\limits_{n=0}^{\infty}\frac{(-1)^n}{n!}(\mathbf{r}'\cdot\nabla)^n G_{ij}(\mathbf{r}).
\end{equation}
Inserting this expression into Eq.~(\ref{Multipole-foundation}) yields
\begin{eqnarray}
	u_i(\mathbf{r}) 
	&={}  &\bigg[F_j - \big( A_{jk} + S_{jk} \big)\nabla_{k}+\frac{1}{2}M_{jkl}\nabla_{k}\nabla_{l}\notag\\ &{}& -\frac{1}{6}N_{jklm}\nabla_k\nabla_l\nabla_m+\dots\bigg]G_{ij}(\mathbf{r}),\label{Multipole-rudimentary}
\end{eqnarray}
with the constant expansion tensors
\begin{eqnarray}
	F_j &={} &\int_{\partial V} \mathrm{d}S' ~f_j(\mathbf{r}'), \label{def_F}\\
	A_{jk} &={} & \int_{\partial V} \mathrm{d}S'~ \frac{1}{2}\Big[f_j(\mathbf{r}') r_k' - f_k(\mathbf{r}') r_j'\Big],	\label{def_A}\\
	S_{jk} &={} & \int_{\partial V} \mathrm{d}S'~  \frac{1}{2}\bigg[f_j(\mathbf{r}') r_k' + f_k(\mathbf{r}') r_j' \bigg], \label{def_S}\\
	M_{jkl} &={} &\int_{\partial V}\mathrm{d}S'~f_j(\mathbf{r}')r_k'r_l',	\label{def_M}	\\
	N_{jklm} &={} &\int_{\partial V}\mathrm{d}S'~f_j(\mathbf{r}')r_k'r_l'r_m'.	\label{def_N}
\end{eqnarray}
Stopping at this order, we here assume that the position $\mathbf{r}$ at which the displacement field is evaluated is sufficiently far away from the surface of the particle, parametrized by $\mathbf{r}'$.

In Eqs.~(\ref{def_F})--(\ref{def_N}),
$\mathbf{F}$ is the total force acting on the rigid particle, resulting, for example, from an external potential. 
This force is transmitted to the surrounding medium and leads to deformation.
In a balanced stationary state, the medium presses back by a counteracting elastic restoring force.
Along the same lines, $\mathbf{\underline{A}}$ is an antisymmetric second-rank tensor that is related to the total torque $\mathbf{T}$ acting on the particle from outside by
\begin{equation}\label{relation_A_T}
	T_l ={} - \epsilon_{ljk} A_{jk},
\end{equation}
with $\epsilon_{jkl}$ denoting the Levi-Civita symbol,
so that
\begin{equation}\label{def_torque}
	\mathbf{T} ={} \int_{\partial V} \mathrm{d}S'~ \mathbf{r}'\times \mathbf{f}(\mathbf{r}').
\end{equation}
The symmetric second-rank tensor $\mathbf{\underline{S}}$ is called stresslet and would be related to possible deformations of the particle surface, if the particle were deformable.
The same applies to the third-rank tensor $\mathbf{\underline{\underline{M}}}$, the fourth-rank tensor $\mathbf{\underline{\underline{\underline{N}}}}$, and
higher-order terms that are not considered here.

\section{Displacement field around a uniformly translated and rotated rigid sphere}
\label{flow_past_sphere}

\begin{figure}
\centerline{\includegraphics[width=.75\columnwidth]{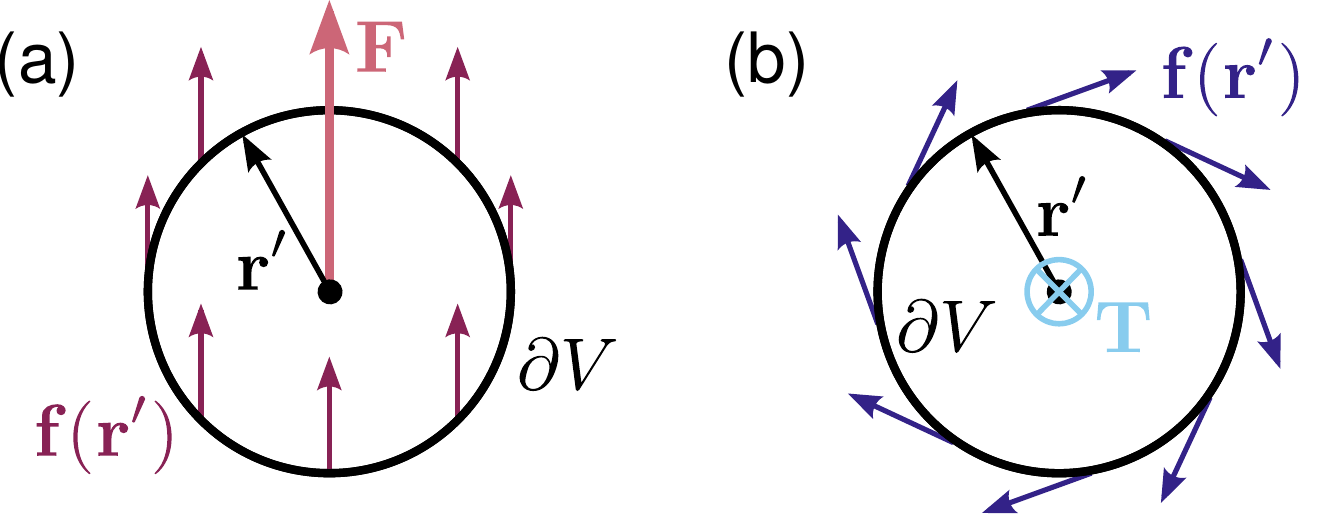}}
\caption{(a) A force $\mathbf{F}$ uniformly translates a spherical shell $\partial V$ (depicted is a central cross-section).
The corresponding force surface density $\mathbf{f}(\mathbf{r}')\sim\mathbf{F}$, represented by the small arrows, is constant on all points $\mathbf{r}'\in\partial V$.
(b) A torque $\mathbf{T}$, attacking at the center of the sphere and normal to the cross-sectional plane, uniformly rotates the sphere. The corresponding force surface density $\mathbf{f}(\mathbf{r}')\sim\mathbf{T}\times\mathbf{r}'$ is tangential to the boundary of the depicted cross section and there of constant magnitude. 
}
\label{fig_force_dens}
\end{figure}

We now consider a rigid spherical particle of radius $a$, centered at $\mathbf{r}=\mathbf{0}$, that is translated and rotated in the elastic medium by a force $\mathbf{F}$ and a torque $\mathbf{T}$, respectively.
The medium shall stick to the particle surface under no-slip boundary conditions.
Moreover, the displacement field $\mathbf{u}(\mathbf{r})$ induced by the sphere shall vanish at infinite distance from the particle.
Thus, the boundary conditions read
\begin{eqnarray}
	\mathbf{u}(|\mathbf{r}|\rightarrow\infty) &={} & \mathbf{0},\qquad
	\mathbf{u}(\mathbf{r}\in\partial V) ={}  \mathbf{U}+\boldsymbol{\Omega}\times\mathbf{r},\quad\label{boundary}
\end{eqnarray}
with $\mathbf{U}$ and $\boldsymbol{\Omega}$ the translation and rotation of the particle, respectively.
Due to the linearity of the Navier--Cauchy equation, we may consider the effect of $\mathbf{F}$ and $\mathbf{T}$ separately and then, at the end, superimpose the solutions.

The integrals in this section are evaluated using \cite{dassios_2012}
\begin{eqnarray}
	\lefteqn{\int_{\partial V} \mathrm{d}S'~\prod\limits_{k=1}^{2N} r'_{i_k}}\notag\\
	 &={} &\frac{4\pi a^{2(N+1)}}{(2N+1)!} \sum\limits_{\substack{k,l,m,n,\ldots=1 \\ \text{all pairwise distinct}}}^{2N}  \delta_{i_k i_l} \delta_{i_m i_n}\dots,\label{polyadic}
\end{eqnarray}
with $N$ denoting the number of dyadics of $\mathbf{r}'$ and all indices $i_k$ ($k=1,\dots,N$) being pairwise distinct.
Only even polyadics of $\mathbf{r}'$ survive the integration over the isotropic spherical surface. 
Therefore, all such integrals lead to combinations of Kronecker deltas, see also Tab.~\ref{table_int}.
Accordingly, the integrations in Eqs.~(\ref{def_F})--(\ref{def_N}) are performed and the results are then inserted back into the right-hand side of Eq.~(\ref{Multipole-rudimentary}).

We start by considering only translations induced by a force $\mathbf{F}$ imposed on the particle. Due to the linearity of the problem, the force surface density $\mathbf{f}(\mathbf{r'})$ ($\mathbf{r'}\in\partial V$) exerted by the sphere onto the medium scales linearly with $\mathbf{F}$, as therefore does the resulting displacement field $\mathbf{u}(\mathbf{r})$. As a consequence, all contributing terms in Eq.~(\ref{Multipole-rudimentary}) must scale linearly in $\mathbf{F}$. Obviously, the factors $\mathbf{f}(\mathbf{r'})$ in the integrals of Eqs.~(\ref{def_F})--(\ref{def_N}) provide this scaling.

Since the solution of the Navier--Cauchy equation under the given boundary conditions is unique, we may continue by an appropriate ansatz for $\mathbf{f}(\mathbf{r'})$. If the result satisfies both the Navier--Cauchy equation and the boundary conditions, we have found the correct solution.

For the translated sphere, displaced in response to a net force $\mathbf{F}$ applied to it, the ansatz corresponds to a uniform force surface density $\mathbf{f}(\mathbf{r'})=\mathbf{F}/4\pi a^2$ ($\mathbf{r'}\in\partial V$) as displayed in Fig.~\ref{fig_force_dens}(a) \cite{dhont1996introduction}. Consequently, all contributions of uneven numbers of gradients in Eq.~(\ref{Multipole-rudimentary}) vanish because of the integration of an uneven number of factors $\mathbf{r'}$ over the surface of the sphere, see, e.g., Eqs.~(\ref{def_A}), (\ref{def_S}), (\ref{def_N}), and our comments above. Moreover, no even orders higher than the one given by $\mathbf{\underline{\underline{M}}}$ may contribute, because the Kronecker deltas resulting via Eq.~(\ref{polyadic}) pairwise combine the gradients in Eq.~(\ref{Multipole-rudimentary}) to Laplacian operators, which via Eq.~(\ref{biharmonic}) leads to vanishing terms. The integral in Eq.~(\ref{def_M}) can then readily be evaluated using Eq.~(\ref{polyadic}) and Tab.~\ref{table_int}. 
As a result, we obtain the familiar form \cite{phanthien1993rigid,puljiz2017forces}
\begin{equation}\label{translating_sphere}
	\mathbf{u}(\mathbf{r}) ={} \left(1+\frac{a^2}{6}\nabla^2\right)\mathbf{\underline{G}}(\mathbf{r})\cdot\mathbf{F}.
\end{equation}
Inserting this expression into the Navier--Cauchy equation and testing the boundary conditions confirms this solution and the chosen ansatz.

We now proceed along the same lines for a torque $\mathbf{T}$ imposed on the particle.
The force density for a uniform rotation leads to the ansatz
$\mathbf{f}(\mathbf{r}')\sim\mathbf{T}\times\mathbf{r}'$ ($\mathbf{r}'\in\partial V$), see also Fig.~\ref{fig_force_dens} (b).
Inserting this $\mathbf{f}(\mathbf{r}')$ into Eqs.~(\ref{def_F})--(\ref{def_N}), only Eq.~(\ref{def_A}) leads to a nonzero result due to antisymmetry.
The expressions in Eq.~(\ref{def_N}) and higher-order terms vanish according to Eqs.~(\ref{biharmonic}) and/or (\ref{rot_condition}) employed in Eq.~(\ref{Multipole-rudimentary}) together with Eq.~(\ref{polyadic}).
The resulting displacement field reads
\begin{equation}\label{rotating_sphere}
	\mathbf{u}(\mathbf{r}) ={}-\frac{1}{2}(\mathbf{T}\times\nabla)\cdot\mathbf{\underline{G}}(\mathbf{r}).
\end{equation}

Adding the two solutions, we obtain the overall displacement field
\begin{equation}\label{isolated_sphere}
	\mathbf{u}(\mathbf{r}) ={} \left(1+\frac{a^2}{6}\nabla^2\right)\mathbf{\underline{G}}(\mathbf{r})\cdot\mathbf{F}-\frac{1}{2}(\mathbf{T}\times\nabla)\cdot\mathbf{\underline{G}}(\mathbf{r}).
\end{equation}
Explicit calculation (see Tab.~\ref{table_dev}) leads to
\begin{eqnarray}
	\mathbf{u}(\mathbf{r})
	 &={} &\frac{1}{16\pi(1-\nu)\mu r}\bigg[ \left(3-4\nu+\frac{1}{3}\left(\frac{a}{r}\right)^2\right)\mathbf{\underline{\hat{I}}} \notag\\ 
	 	 	 	 &{}&
	 	 	 	 + 
	 \left(1-\left(\frac{a}{r}\right)^2\right)
	 \mathbf{\hat{r}}\mathbf{\hat{r}} \bigg]\cdot\mathbf{F} 
	 + \frac{1}{8\pi\mu}\mathbf{T}\times\frac{\mathbf{r}}{r^3}. \label{isolated_sphere_explicit}
\end{eqnarray}
Following the boundary conditions in Eq.~(\ref{boundary}), we find
\begin{eqnarray}\label{stokes_law}
	\mathbf{U} ={} \frac{5-6\nu}{24\pi(1-\nu)\mu a}\mathbf{F}, \quad
	\boldsymbol{\Omega} ={} \frac{1}{8\pi\mu a^3}\mathbf{T}.
\end{eqnarray}
\begin{table*}
	    \begin{eqnarray*}
	     	\int_{\partial V} \mathrm{d}S'  &={} &4\pi a^2\\
	     	\int_{\partial V} \mathrm{d}S'~r'_j r'_k &={}&  \frac{4\pi a^4}{3}\delta_{jk} \notag\\
	     	\int_{\partial V} \mathrm{d}S'~r'_j r'_k r'_l r'_m &={} &\frac{4\pi a^6}{15}\left[\delta_{jk}\delta_{lm} + \delta_{jl}\delta_{km} + \delta_{jm}\delta_{kl}\right]\\
	     	\int_{\partial V} \mathrm{d}S'~r'_j r'_k r'_l r'_m r'_n r'_o &={} &\frac{4\pi a^8}{105}[
	     	\delta_{jk}(\delta_{lm}\delta_{no}+\delta_{ln}\delta_{mo}+\delta_{lo}\delta_{mn})
	     	\\
	     	&{}& 
	     	+\delta_{jl}
	     	(\delta_{km}\delta_{no}+\delta_{kn}\delta_{mo}+\delta_{ko}\delta_{mn})
	     	\\
	     	&{}& 
	     	+\delta_{jm}
	     	(\delta_{kl}\delta_{no}+\delta_{kn}\delta_{lo}+\delta_{ko}\delta_{ln})
	     	\\
	     	&{}& 
	     	+\delta_{jn}
	     	(\delta_{kl}\delta_{mo}+\delta_{km}\delta_{lo}+\delta_{ko}\delta_{lm})
	     	\\
	     	&{}& 
	     	+\delta_{jo}
	     	(\delta_{kl}\delta_{mn}+\delta_{km}\delta_{ln}+\delta_{kn}\delta_{lm})]\\
	     	\int_{\partial V}\mathrm{d}S' r_j'r_k'r_l'r_m'r_n'r_o'r_p'r_q' &={}& \frac{4\pi a^{10}}{945}[\delta_{jk}\delta_{lm}\delta_{no}\delta_{pq}+\text{all permutations}]
	     \end{eqnarray*}
	\caption{
		Equation (\ref{polyadic}) for $N=0,1,2,3$ (from top to bottom).
	}
	\label{table_int}
\end{table*}
\begin{table*}
	    \begin{eqnarray*}
	     	16\pi(1-\nu)\mu\,\nabla_k G_{ij}  &={} &	\frac{1}{r^3}[-(3-4\nu)r_k\delta_{ij}+r_j\delta_{ik}+r_i\delta_{jk}]-\frac{3}{r^5}r_i r_j r_k	\\
	     	16\pi(1-\nu)\mu\,\nabla^2 G_{ij} &={} & \frac{2}{r^3}\delta_{ij}-\frac{6}{r^5}r_i r_j
	     \end{eqnarray*}
	\caption{
		Derivatives of the Green's tensor, see Eq.~(\ref{greens_function}).
	}
	\label{table_dev}
\end{table*}

\section{Fax\'en relations}
\label{faxen}

Now we consider the opposite situation, in which a given displacement field $\mathbf{u}_0(\mathbf{r})$ in the elastic medium translates and rotates a rigid spherical particle centered at the origin.
Then, we can reformulate the second expression in Eq.~(\ref{boundary}) as
\begin{equation}
	 \mathbf{U} + \boldsymbol{\Omega}\times\mathbf{r}
	 ={} \mathbf{u}_0(\mathbf{r}) + \int_{\partial V} \mathrm{d}S'~\mathbf{\underline{G}}(\mathbf{r}-\mathbf{r}')\cdot\mathbf{f}(\mathbf{r}'),	\label{faxen_multipole}
\end{equation}
which must hold true for all $\mathbf{r}\in\partial V$.
The left-hand side corresponds to the translations and rotations of the points on the spherical surface shell, see Eq.~(\ref{boundary}).
Due to stick boundary conditions, these must be equal to the displacement field of the surrounding elastic medium at all surface points $\mathbf{r}$, see the right-hand side of Eq.~(\ref{faxen_multipole}).
However, the given displacement field $\mathbf{u}_0(\mathbf{r})$ will in general not be of a constant or antisymmetrically distributed form over the whole surface.
Thus, it would in general imply a deformation of the spherical surface shell.
This conflicts with the assumed rigidity of the particle.
Therefore, the contributions in $\mathbf{u}_0(\mathbf{r})$ that would lead to such impossible deformations will be balanced by counteracting effects by the particle itself.
The displacement fields resulting from these counteracting effects are represented by the additional integral term in Eq.~(\ref{faxen_multipole}).
They are given by all terms in Eq.~(\ref{Multipole-rudimentary}) that do not contribute to Eq.~(\ref{isolated_sphere}), i.e., by an infinite series.
Multiplying Eq.~(\ref{faxen_multipole}) by polyadics of $\mathbf{r}$ and performing another integral over the surface of the sphere, the expansion tensors in Eqs.~(\ref{def_F})--(\ref{def_N}) appear from the integral term.
Using the procedure described below, explicit expressions are calculated for the third-rank tensor $\mathbf{\underline{\underline{M}}}$ and the fourth-rank tensor $\mathbf{\underline{\underline{\underline{N}}}}$.
Expressions for $\mathbf{U}$, $\boldsymbol{\Omega}$, and $\mathbf{\underline{S}}$ were already calculated in Ref.~\onlinecite{puljiz2017forces} and read
\begin{eqnarray}
	\mathbf{U} &={}&
\frac{5-6\nu}{24\pi(1-\nu)\mu a}\mathbf{F}+ \left(1+\frac{a^2}{6}\nabla^2\right)\mathbf{u}_0(\mathbf{r})\bigg|_{\mathbf{r}=\mathbf{0}}, \label{faxen_F}
	\\
	\boldsymbol{\Omega} &={}&
		\frac{1}{8\pi\mu a^3}\mathbf{T}+
		 \frac{1}{2}\nabla\times\mathbf{u}_0(\mathbf{r})|_{\mathbf{r}=\mathbf{0}}\label{faxen_O},
	\\
	\mathbf{\underline{S}} &={}& -\frac{4\pi(1-\nu)\mu a^3}{4-5\nu}\left(1+\frac{a^2}{10}\nabla^2\right)\bigg[
		\frac{1}{1-2\nu}\mathbf{\underline{\hat{I}}}\,\nabla\cdot\mathbf{u}_0(\mathbf{r})
		\notag\\
	&{}&+\frac{5}{2}\Big(
		\nabla \mathbf{u}_{0}(\mathbf{r})
		+\big(\nabla \mathbf{u}_{0}(\mathbf{r})\big)^\text{T}
	\Big)
	\bigg]\bigg|_{\mathbf{r}=\mathbf{0}},	\label{S_jk}
\end{eqnarray}
respectively ($^\text{T}$ denotes the transpose).

In order to calculate $\mathbf{\underline{\underline{M}}}$ as defined in Eq.~(\ref{def_M}), we insert the series expansion of  $\mathbf{u}_0(\mathbf{r})$ around the origin (where the sphere is centered),
\begin{eqnarray}
	u_{0,i}(\mathbf{r}) &={} &u_{0,i}(\mathbf{r}=\mathbf{0}) + r_p [\nabla_p u_{0,i}(\mathbf{r})]_{\mathbf{r}=\mathbf{0}} \notag\\ &{}&+ \frac{1}{2!}r_pr_q [\nabla_p\nabla_q u_{0,i}(\mathbf{r})]_{\mathbf{r}=\mathbf{0}} \notag\\
	& {} & +\frac{1}{3!}r_pr_qr_r [\nabla_p\nabla_q\nabla_r u_{0,i}(\mathbf{r})]_{\mathbf{r}=\mathbf{0}} + ..., \label{taylor_u}
\end{eqnarray}
into Eq.~(\ref{faxen_multipole}) and substitute the expression given in Eq.~(\ref{faxen_F}) into $\mathbf{U}$ in Eq.~(\ref{faxen_multipole}).
Then integrating over $\int_{\partial V} \mathrm{d}S~r_k r_l$, we obtain
\begin{eqnarray} 
	0 &={} & \frac{4\pi a^6}{15}\left(1+\frac{a^2}{14}\nabla^2\right)\left[\nabla_k\nabla_l-\frac{1}{3}\delta_{kl}\nabla^2\right]u_{0,i}(\mathbf{r})\bigg|_{\mathbf{r}=\mathbf{0}}	\notag\\
	&{}&-\frac{(5-6\nu)a^3}{18(1-\nu)\mu}F_i\delta_{kl}\notag\\
	&{}& +\int_{\partial V}\mathrm{d}S'\int_{\partial V}\mathrm{d}S~r_kr_l\, G_{ij}(\mathbf{r}-\mathbf{r}')f_j(\mathbf{r}'). \label{boundary_mjkl}
\end{eqnarray}
Evaluation of the integral on the right-hand side yields (see Appendix A)
\begin{eqnarray}
	\lefteqn{\int_{\partial V}\mathrm{d}S'\int_{\partial V}\mathrm{d}S~r_kr_l\, G_{ij}(\mathbf{r}-\mathbf{r}')f_j(\mathbf{r}')} \notag\\
	&={}&
	\frac{a}{210(1-\nu)\mu}\bigg\{
	3(13-14\nu) M_{ikl} + 4 M_{nni}\delta_{kl} \notag\\
					&{}&\qquad\quad- 3(M_{kil}+M_{lki}+M_{nnl}\delta_{ik}+M_{nnk}\delta_{il})
	\bigg\}\notag\\
	&{}&
	+\frac{a^3}{105(1-\nu)\mu}\left\{
	 2(11-14\nu)F_i\delta_{kl}+F_k \delta_{il} + F_{l} \delta_{ik}
	\right\}
	,\qquad\label{int_m_jkl}
\end{eqnarray}
noting that $M_{ikl}=M_{ilk}$, see Eq.~(\ref{def_M}).
Reinserting Eq.~(\ref{int_m_jkl}) into Eq.~(\ref{boundary_mjkl}), we obtain
\begin{eqnarray}
	0 &={}& \Gamma_{jkl} + 3(13-14\nu) M_{jkl} + 4 M_{nnj}\delta_{kl} \notag\\
				&{}&- 3(M_{kjl}+M_{ljk}+M_{nnl}\delta_{jk}+M_{nnk}\delta_{jl}),	\label{boundary_mjkl_2}
\end{eqnarray}
with
\begin{eqnarray}
	\Gamma_{jkl}
	 &={} & 
	 2a^2\left[
	 	 	F_k\delta_{jl} + F_l\delta_{jk} -\frac{(43-42\nu)}{6}F_j\delta_{kl}
	 	 \right]\notag\\
	 	 &{}&
	 +56(1-\nu)\pi\mu a^5\left(1+\frac{a^2}{14}\nabla^2\right)\Bigg[\nabla_k\nabla_l\notag\\
	 &{}&-\frac{1}{3}\delta_{kl}\nabla^2\Bigg]u_{0,j}(\mathbf{r})\bigg|_{\mathbf{r}=\mathbf{0}}.	\label{Gamma}
\end{eqnarray}

Equation (\ref{boundary_mjkl_2}) is symmetric with respect to $k\leftrightarrow l$, but not with respect to $j\leftrightarrow k$ or $j\leftrightarrow l$.
To solve Eq.~(\ref{boundary_mjkl_2}) for $M_{jkl}$, it is useful to decompose it into a symmetric and a partially antisymmetric part.
We define
\begin{eqnarray}
	M_{jkl}^\text{S} &={}& M_{jkl} + M_{kjl} + M_{lkj},\label{M_jkl_S}\\
	M_{jkl}^\text{A} &={}& M_{jkl} - M_{kjl} - M_{lkj},\label{M_jkl_A}
\end{eqnarray}
so that $M_{jkl} ={} (M_{jkl}^\text{S}+M_{jkl}^\text{A})/2$.

Symmetrizing and antisymmetrizing Eq.~(\ref{boundary_mjkl_2}) according to Eqs.~(\ref{M_jkl_S}) and (\ref{M_jkl_A}), respectively, we obtain
\begin{widetext}
\noindent
\begin{eqnarray}
	M_{jkl}^\text{S} 
	&={} &-\frac{1}{33-42\nu}\bigg\{
	\Gamma_{jkl}^\text{S} - 2(M_{nnj}\delta_{kl}+M_{nnk}\delta_{jl}
	+M_{nnl}\delta_{jk})
	\bigg\},	\label{M_jkl_S_explicit}	\\
	M_{jkl}^\text{A} &={} &-\frac{1}{42(1-\nu)}\bigg\{
	3M_{jkl}^\text{S}
	+\Gamma_{jkl}^\text{A}
	+10 M_{nnj}\delta_{kl} -4(M_{nnk}\delta_{jl}+M_{nnl}\delta_{jk})\bigg\}.	\label{M_jkl_A_explicit}
\end{eqnarray}
Thus, solving Eq.~(\ref{boundary_mjkl_2}) for $M_{jkl}$, from $(M_{jkl}^\text{S}+M_{jkl}^\text{A})/2$ we obtain
\begin{eqnarray}
	M_{jkl}
	&={}&
	-\frac{1}{84(1-\nu)(11-14\nu)}\bigg\{
	(13-14\nu)\Gamma_{jkl}^\text{S}+(11-14\nu)\Gamma_{jkl}^\text{A}
	\notag\\
	&{}&
	+28(3-4\nu)
		M_{nnj}\delta_{kl}-14(5-6\nu)(M_{nnk}\delta_{jl} + M_{nnl}\delta_{jk})
	\bigg\}.\label{M_jkl_formal}
\end{eqnarray}
Setting and summing over $k=l$ in Eq.~(\ref{M_jkl_formal}), we find the remaining unknown,
\begin{equation}\label{M_nnj}
	M_{nnj} = \frac{a^2}{3}F_j + \frac{4\pi(1-\nu)\mu a^5}{2-3\nu}\bigg\{\frac{1}{3}\nabla^2 u_{0,j}(\mathbf{r})-\nabla_j\nabla\cdot\mathbf{u}_0(\mathbf{r})\bigg\}\bigg|_{\mathbf{r}=\mathbf{0}}.
\end{equation}
This finally leads us to
\begin{eqnarray}
	M_{jkl}
		&={}&
			\frac{a^2}{3}F_j \delta_{kl}
			-\frac{2\pi \mu a^5}{3(11-14\nu)}
				\left(
				1+\frac{a^2}{14}\nabla^2
				\right)
			\bigg[
					2\Big(2(6-7\nu)\nabla_k\nabla_l u_{0,j}(\mathbf{r}) + \nabla_j\nabla_k u_{0,l}(\mathbf{r})+ \nabla_j\nabla_l u_{0,k}(\mathbf{r})\Big)
					\notag\\
							&{}&
					-\frac{1}{2-3\nu}\nabla^2 \Big(	2(7-20\nu+14\nu^2)u_{0,j}(\mathbf{r})\delta_{kl}+(3-4\nu)(u_{0,k}(\mathbf{r})\delta_{jl}+u_{0,l}(\mathbf{r})\delta_{jk})\Big)	\notag\\
					&{}&-\frac{1}{2-3\nu}\Big(
						2(3-4\nu)\delta_{kl}\nabla_j
						-(5-6\nu)(\delta_{jl}\nabla_k+\delta_{jk}\nabla_l) 
					\Big)\nabla\cdot\mathbf{u}_0(\mathbf{r})
					\bigg]\bigg|_{\mathbf{r}=\mathbf{0}}
					.\qquad
		\label{M_jkl_explicit}
\end{eqnarray}
The corresponding incompressible-creeping-flow hydrodynamic result is obtained by setting $\nabla\cdot\mathbf{u}_0(\mathbf{r})=0$ as well as $\nu=1/2$, and by replacing the shear modulus $\mu$ by the viscosity $\eta$,
\begin{eqnarray}
	M_{jkl}^\text{h} &={} & 	\frac{a^2}{3}F_j \delta_{kl}-\frac{\pi\eta a^5}{3}\left(1+\frac{a^2}{14}\nabla^2\right)\bigg[
		5\nabla_k\nabla_l u_{0,j}(\mathbf{r})
		+\nabla_j\nabla_k u_{0,l}(\mathbf{r})
		+\nabla_j\nabla_l u_{0,k}(\mathbf{r})
		\notag\\
		&{}&
		- \nabla^2\Big(
		u_{0,j}(\mathbf{r})\delta_{kl}
		+u_{0,k}(\mathbf{r})\delta_{jl}
		+u_{0,l}(\mathbf{r})\delta_{jk}
		\Big)
	\bigg]\bigg|_{\mathbf{r}=\mathbf{0}},\qquad
\end{eqnarray}
where $\mathbf{u}_0(\mathbf{r})$ is interpreted as the flow field.

To obtain the fourth-rank tensor $\mathbf{\underline{\underline{\underline{N}}}}$, we substitute the expression from Eq.~(\ref{faxen_O}) into Eq.~(\ref{faxen_multipole}), use Eq.~(\ref{rot_condition}), insert Eq.~(\ref{taylor_u}), and integrate over $\int_ {\partial V}\mathrm{d}S~r_k r_l r_m$, yielding (see Appendix B)
\begin{eqnarray}
	0 &={}& -(17-18\nu)N_{jklm}  + N_{kjlm} + N_{ljkm} + N_{mjkl}	+N_{nnkm}\delta_{jl} + N_{nnlm}\delta_{jk} + N_{nnkl}\delta_{jm}\notag\\
	&{}&-\frac{4}{5}(
		N_{nnjk}\delta_{lm} + N_{nnjl}\delta_{km} + N_{nnjm}\delta_{kl}
	)
	-\Gamma_{jklm},
	\label{N_eq}
\end{eqnarray}
with
\begin{eqnarray}
	\Gamma_{jklm} &={}&
	\frac{4\pi(1-\nu)\mu a^5}{4-5\nu}\left(1+\frac{a^2}{10}\nabla^2\right)\Bigg\{
	\frac{4(3-4\nu)}{5(1-2\nu)}\nabla\cdot\mathbf{u}_0(\mathbf{r})(\delta_{jk}\delta_{lm}+\delta_{jl}\delta_{km}+\delta_{jm}\delta_{kl})\notag\\
	&{}&-2\big[
		\delta_{jk}(\nabla_l u_{0,m}(\mathbf{r})+\nabla_m u_{0,l}(\mathbf{r}))
		+ \delta_{jl}(\nabla_k u_{0,m}(\mathbf{r})+\nabla_m u_{0,k}(\mathbf{r}))
		+\delta_{jm}(\nabla_k u_{0,l}(\mathbf{r})+\nabla_l u_{0,k}(\mathbf{r}))
	\big]\notag\\
	&{}&
	+\frac{44-45\nu}{5}\big[
			\delta_{kl}(\nabla_j u_{0,m}(\mathbf{r})+\nabla_m u_{0,j}(\mathbf{r}))
			+ \delta_{km}(\nabla_j u_{0,l}(\mathbf{r})+\nabla_l u_{0,j}(\mathbf{r}))
			+\delta_{lm}(\nabla_j u_{0,k}(\mathbf{r})+\nabla_k u_{0,j}(\mathbf{r}))
		\big]
	\Bigg\}\Bigg|_{\mathbf{r}=\mathbf{0}}\notag\\
	&{}&
		+ \frac{24\pi(1-\nu)\mu a^7}{5}\left(1+\frac{a^2}{18}\nabla^2\right)\left\{
			\nabla_k\nabla_l\nabla_m - \frac{1}{5}(\delta_{kl}\delta_{mn}+\delta_{km}\delta_{ln}+\delta_{kn}\delta_{lm})\nabla^2 \nabla_n
		\right\}u_{0,j}(\mathbf{r})\bigg|_{\mathbf{r}=\mathbf{0}}
		\notag\\
	&{}&
	 +\frac{9(1-\nu)a^2}{5}\big[\delta_{kl}\epsilon_{jmn}+\delta_{km}\epsilon_{jln}+\delta_{lm}\epsilon_{jkn}\big]T_n.
\end{eqnarray}
The formal solution of Eq.~(\ref{N_eq}) reads
\begin{eqnarray}
	N_{jklm} &={}& -\frac{1}{36(1-\nu)(7-9\nu)}\left\{
	3(5-6\nu)\Gamma_{jklm}+\Gamma_{kjlm}+\Gamma_{ljkm}+\Gamma_{mjkl}
	\right\}\notag\\
	&{}& + \frac{1}{10(1-\nu)(7-9\nu)}\{
		(4-5\nu)[
		 N_{nnlm}\delta_{jk}+ N_{nnkm}\delta_{jl} + N_{nnkl}\delta_{jm}
		]	\notag\\
		&{}&-(3-4\nu)[
			N_{nnjk}\delta_{lm}
			+N_{nnjl}\delta_{km}
			+N_{nnjm}\delta_{kl} 
		]
	\}	\label{N_formal}
\end{eqnarray}
and can be obtained by manipulating the whole equation in agreement with the definitions
\begin{eqnarray}
	N_{jklm}^\text{S} &={} & N_{jklm} + N_{kjlm} + N_{ljkm} + N_{mjkl},\\
	N_{jklm}^\text{A} &={} & N_{jklm} - N_{kjlm} - N_{ljkm} - N_{mjkl},
\end{eqnarray}
so that $N_{jklm} ={} (N_{jklm}^\text{S}+N_{jklm}^\text{A})/2$, similarly to the procedure for $M_{jkl}$.
$N_{nnlm}$ is obtained by setting and summing over $j=k$ in Eq.~(\ref{N_formal}),
\begin{equation}
	N_{nnlm} ={} -\frac{5}{18(7-10\nu)(8-9\nu)}\{
		2(8-9\nu)\Gamma_{nnlm}+\Gamma_{lmnn} + \Gamma_{mlnn}\} - \frac{3-4\nu}{(7-10\nu)(8-9\nu)}N_{nnoo}\delta_{lm}.	\label{N_nnlm}
\end{equation}
Moreover, from the trace of Eq.~(\ref{N_nnlm}),
\begin{equation}
	N_{nnoo} ={} -\frac{1}{13-18\nu}\Gamma_{nnoo} ={} -\frac{8\pi(1-\nu)\mu a^5}{1-2\nu}\nabla\cdot\mathbf{u}_0(\mathbf{r})|_{\mathbf{r}=\mathbf{0}} ={} a^2S_{nn}
\end{equation}
is found.
Finally,
\begin{eqnarray}
	N_{jklm}
		&={}&
			-\frac{a^2}{10}\big[\delta_{kl}\epsilon_{jmn}+\delta_{km}\epsilon_{jln}+\delta_{lm}\epsilon_{jkn}\big]T_n\notag\\
			&{}&
			-\frac{4\pi(1-\nu)\mu a^5}{5(4-5\nu)}\left(1+\frac{a^2}{10}\nabla^2\right)\bigg[
				\frac{1}{1-2\nu}\nabla\cdot\mathbf{u}_0(\mathbf{r})(\delta_{jk}\delta_{lm}+\delta_{jl}\delta_{km}+\delta_{jm}\delta_{kl})\notag\\
				&{}&
				+\frac{5}{2}\big[
							\delta_{kl}(\nabla_j u_{0,m}(\mathbf{r})+\nabla_m u_{0,j}(\mathbf{r}))
							+ \delta_{km}(\nabla_j u_{0,l}(\mathbf{r})+\nabla_l u_{0,j}(\mathbf{r}))
							+\delta_{lm}(\nabla_j u_{0,k}(\mathbf{r})+\nabla_k u_{0,j}(\mathbf{r}))
						\big]
			\bigg]\bigg|_{\mathbf{r}=\mathbf{0}}\notag\\
			&{}&+B_{jklm},	\label{N_jklm}
\end{eqnarray}
with
\begin{eqnarray}
	B_{jklm} &={} &
	-\frac{2\pi\mu a^7}{15(7-9\nu)}\left(1+\frac{a^2}{18}\nabla^2\right)\Bigg\{
					3(5-6\nu)
						\nabla_k\nabla_l\nabla_m u_{0,j}(\mathbf{r})
					+\nabla_j\nabla_k\nabla_l u_{0,m}(\mathbf{r})
					+\nabla_j\nabla_l\nabla_m u_{0,k}(\mathbf{r})\notag\\
										&{}&
					+\nabla_j\nabla_k\nabla_m u_{0,l}(\mathbf{r})
			-\frac{2(7-9\nu)}{5}\nabla^2 (
				\delta_{kl}\nabla_m + \delta_{km}\nabla_l + \delta_{lm}\nabla_k
			) u_{0,j}(\mathbf{r})\notag\\
			&{}&
			-\frac{1-2\nu}{5(7-10\nu)}\nabla^2
			\big[
							\delta_{kl}(\nabla_j u_{0,m}(\mathbf{r})+\nabla_m u_{0,j}(\mathbf{r}))
							+ \delta_{km}(\nabla_j u_{0,l}(\mathbf{r})+\nabla_l u_{0,j}(\mathbf{r}))
							+\delta_{lm}(\nabla_j u_{0,k}(\mathbf{r})+\nabla_k u_{0,j}(\mathbf{r}))
						\big]
			\notag\\
			&{}&
			-\frac{(3-4\nu)}{7-10\nu}\nabla^2
		\big[
					\delta_{jk}(\nabla_l u_{0,m}(\mathbf{r})+\nabla_m u_{0,l}(\mathbf{r}))
					+ \delta_{jl}(\nabla_k u_{0,m}(\mathbf{r})+\nabla_m u_{0,k}(\mathbf{r}))
					+\delta_{jm}(\nabla_k u_{0,l}(\mathbf{r})+\nabla_l u_{0,k}(\mathbf{r}))
								\big]
			\notag\\
			&{}&
		+\frac{2}{7-10\nu}\big[(4-5\nu)(\delta_{jk}\nabla_l\nabla_m
		+\delta_{jl}\nabla_k\nabla_m
		+\delta_{jm}\nabla_k\nabla_l
		)\notag\\
							&{}&
		-(3-4\nu)(\delta_{kl}\nabla_j\nabla_m
				+\delta_{km}\nabla_j\nabla_l
				+\delta_{lm}\nabla_j\nabla_k
				)
		\big]\nabla\cdot\mathbf{u}_0(\mathbf{r})
				\Bigg\}\Bigg|_{\mathbf{r}=\mathbf{0}}.
\end{eqnarray}
The corresponding result in incompressible low-Reynolds-number hydrodynamics is obtained for $\nabla\cdot\mathbf{u}_0(\mathbf{r})=0$, then setting $\nu=1/2$, and replacing $\mu$ by $\eta$. It reads
\begin{eqnarray}
	N_{jklm}^\text{h} &={}& -\frac{a^2}{10}\big[\delta_{kl}\epsilon_{jmn}+\delta_{km}\epsilon_{jln}+\delta_{lm}\epsilon_{jkn}\big]T_n\notag\\
	&{}&
				-\frac{2\pi\eta a^5}{3}\bigg[
					\big[
								\delta_{kl}(\nabla_j u_{0,m}(\mathbf{r})+\nabla_m u_{0,j}(\mathbf{r}))
								+ \delta_{km}(\nabla_j u_{0,l}(\mathbf{r})+\nabla_l u_{0,j}(\mathbf{r}))
								+\delta_{lm}(\nabla_j u_{0,k}(\mathbf{r})+\nabla_k u_{0,j}(\mathbf{r}))
							\big]
				\bigg]\bigg|_{\mathbf{r}=\mathbf{0}} \notag\\
				&{}&+ B_{jklm}^\text{h},
\end{eqnarray}
with
\begin{eqnarray}
	B_{jklm}^\text{h} &={} &
	-\frac{4\pi\eta a^7}{75}\left(1+\frac{a^2}{18}\nabla^2\right)\Bigg\{
				6\,
						\nabla_k\nabla_l\nabla_m u_{0,j}(\mathbf{r})
					+\nabla_j\nabla_k\nabla_l u_{0,m}(\mathbf{r})
					+\nabla_j\nabla_l\nabla_m u_{0,k}(\mathbf{r})
					+\nabla_j\nabla_k\nabla_m u_{0,l}(\mathbf{r})\notag\\
					&{}&
			-\nabla^2 (
				\delta_{kl}\nabla_m + \delta_{km}\nabla_l + \delta_{lm}\nabla_k
			) u_{0,j}(\mathbf{r})
			-\frac{1}{2}\nabla^2
		\big[
					\delta_{jk}(\nabla_l u_{0,m}(\mathbf{r})+\nabla_m u_{0,l}(\mathbf{r}))
				\notag\\
							&{}&
					+ \delta_{jl}(\nabla_k u_{0,l}(\mathbf{r})+\nabla_l u_{0,k}(\mathbf{r}))
					+\delta_{jm}(\nabla_k u_{0,l}(\mathbf{r})+\nabla_l u_{0,k}(\mathbf{r}))
								\big]
				\Bigg\}\Bigg|_{\mathbf{r}=\mathbf{0}}.
\end{eqnarray}

At this point, we add some brief remarks concerning our results.
The $\mathbf{\underline{\underline{M}}}$-tensor in Eq.~(\ref{M_jkl_explicit}) for setting and summing over $k=l$ becomes $M_{jnn} = a^2 F_j$, which is in agreement with the definition in Eq.~(\ref{def_M}).
Therefore, $\mathbf{\underline{\underline{M}}}$ consists of two contributions: one resulting from the active displacement of the sphere by an external force $\mathbf{F}$ included by $M_{jnn}$; 
the other because of the rigidity of the sphere, which leads to a passive resistance to deformations that would be induced by the imposed matrix deformation $\mathbf{u}_0(\mathbf{r})$.
It is therefore natural to split up the corresponding term in Eq.~(\ref{Multipole-rudimentary}) accordingly, i.e.,
\begin{equation}\label{M_split}
	\frac{1}{2}M_{jkl}\nabla_k\nabla_lG_{ij}(\mathbf{r}) ={} \frac{a^2}{6}F_j\nabla^2 G_{ij}(\mathbf{r}) + \frac{1}{2}\left(M_{jkl}-\frac{1}{3}M_{jnn}\delta_{kl}\right)\nabla_k\nabla_l G_{ij}(\mathbf{r}).
\end{equation}
Together with the other force term in Eq.~(\ref{Multipole-rudimentary}), $F_j G_{ij}(\mathbf{r})$, and in the absence of $\mathbf{u}_0(\mathbf{r})$, the first term on the right-hand side of Eq.~(\ref{M_split}) brings us back to Eq.~(\ref{translating_sphere}) and confirms the result in Sec.~\ref{flow_past_sphere} from another angle (the same applies to the hydrodynamic case).

The $\mathbf{\underline{\underline{\underline{N}}}}$-tensor in Eq.~(\ref{N_jklm}) can be written in the more legible form
\begin{equation}\label{N_short}
	N_{jklm} ={} \frac{a^2}{5} (\delta_{kl}\delta_{mn}+\delta_{km}\delta_{ln}+\delta_{kn}\delta_{lm})[A_{jn} + S_{jn}] + B_{jklm}.
\end{equation}
Since $B_{jknn} = 0$ (and also $B_{jknn}^\text{h}=0$ in the hydrodynamic case), the correct contraction $N_{jknn} ={} a^2[A_{jk}+S_{jk}]$
is obtained in accordance with the definition of $\mathbf{\underline{\underline{\underline{N}}}}$ in Eq.~(\ref{def_N}).
Inserting Eq.~(\ref{N_short}) into the corresponding term on the right-hand side of Eq.~(\ref{Multipole-rudimentary}), we thus find the identity
\begin{equation}\label{N_identity}
	-\frac{1}{6}N_{jklm}\nabla_k\nabla_l\nabla_mG_{ij}(\mathbf{r}) ={} -\frac{a^2}{10}S_{jk}\nabla^2\nabla_k G_{ij}(\mathbf{r})
	- \frac{1}{6}B_{jklm}\nabla_k\nabla_l\nabla_mG_{ij}(\mathbf{r}),
\end{equation}
where the antisymmetric contribution from $A_{jk}$ vanishes due to Eqs.~(\ref{rot_condition}) and (\ref{relation_A_T}).

As a rule, expansion tensors of odd rank always contain the external force $\mathbf{F}$, whereas the tensors of even rank always contain the external torque $\mathbf{T}$.
However, from rank four on, these contributions to the resulting displacement and rotation of the sphere vanish, if inserted into Eq.~(\ref{Multipole-rudimentary}), because of Eqs.~(\ref{laplace_div_u})--(\ref{rot_condition}). 
For instance, in the rank-five tensor beyond $\mathbf{\underline{\underline{\underline{N}}}}$ in Eq.~(\ref{Multipole-rudimentary}), the external force may appear in the form $\sim \mathbf{F}\mathbf{\hat{\underline{I}}}\,\mathbf{\hat{\underline{I}}}$ and permutations thereof.
Upon contraction with $\nabla\nabla\nabla\nabla\mathbf{\underline{G}}(\mathbf{r})$ the next-higher-order contribution in Eq.~(\ref{Multipole-rudimentary}) either contains $\nabla^2\nabla\cdot\mathbf{\underline{G}}(\mathbf{r})=\mathbf{0}$ or $\nabla^4\mathbf{\underline{G}}(\mathbf{r})=\mathbf{\underline{0}}$.
Therefore, all expansion tensors of higher rank than $\mathbf{\underline{\underline{M}}}$ only contribute to the rigidity-induced resistance of the sphere to deformations.


Altogether, in differential form, the total displacement field around the sphere reads
\begin{eqnarray}
	u_i(\mathbf{r})
	&={}&  \Bigg[F_j \left(1 + \frac{a^2}{6}\nabla^2\right)
	+ \frac{1}{2}\epsilon_{jkl}T_l\nabla_{k} \Bigg]G_{ij}(\mathbf{r})
	\notag\\
	&{}&-\Bigg[ S_{jk}\left(1+\frac{a^2}{10}\nabla^2\right)\nabla_k
	-\frac{1}{2}\left(M_{jkl}-\frac{1}{3}M_{jnn}\delta_{kl}\right)\nabla_{k}\nabla_{l}
	+ \frac{1}{6}B_{jklm}\nabla_k\nabla_l\nabla_m
	-\dots\Bigg]G_{ij}(\mathbf{r}),
	\label{eq54}
\end{eqnarray}
which is correct up to order $(\nabla^n \mathbf{u}_0(\mathbf{r})|_{\mathbf{r}=\mathbf{0}})r^{-m}$, with $n,m\in\mathbb{N}$ and $n+m\le 6$.
The first square bracket expresses the exact displacement field that is created by the rigid translation and rotation of the sphere due to an external force $\mathbf{F}$ and torque $\mathbf{T}$, respectively \cite{puljiz2017forces}.
This here directly follows from the explicit analytical expressions for the tensors $\mathbf{\underline{\underline{M}}}$ and $\mathbf{\underline{\underline{\underline{N}}}}$ as derived above.
The second square bracket contains the displacement field induced by the rigidity of the sphere in resistance to deformations from any other distortion of the embedding elastic matrix given by $\mathbf{u}_0(\mathbf{r})$.
These conclusions directly apply to incompressible low-Reynolds-number hydrodynamics as well.
\end{widetext}

\section{Matrix-mediated interactions between spherical inclusions}
\label{matrixmediated}

We now consider a system consisting of $N$ identical rigid spheres of radius $a$ centered at positions $\mathbf{r}_i$, $i=1,2,\dots,N$, embedded under no-slip surface conditions in the linearly elastic matrix.
If external forces and/or torques are applied to the particles, these forces and/or torques are transmitted to the elastic environment, there leading to deformations, which in turn results in matrix-mediated interactions between the particles \cite{puljiz2017forces}.
These interactions can be calculated using the so-called
method of reflections.
In principle, the method proceeds by an expansion in the inverse interparticle distance.
So far, the matrix-mediated interactions have been calculated up to (including) fourth order for a compressible linearly elastic matrix \cite{puljiz2017forces}.
With the now-available $\mathbf{\underline{\underline{M}}}$- and $\mathbf{\underline{\underline{\underline{N}}}}$-tensors, we continue this iteration scheme up to (including) the sixth order in the inverse interparticle separation.
We have described the principles of the underlying scheme in detail in Ref.~\onlinecite{puljiz2017forces}.
Again, for incompressible systems, an analogon exists for low-Reynolds-number hydrodynamics.

To proceed along these lines, the displacement field created by the force $\mathbf{F}_i$ and/or the torque $\mathbf{T}_i$ on the $i$th spherical inclusion in the absence of any other sphere is denoted as $\mathbf{u}_i^{(0)}(\mathbf{r})$. 
The resulting translation and rotation of the sphere is referred to as $\mathbf{U}_i^{(0)}$ and $\boldsymbol{\Omega}_i^{(0)}$, respectively, which according to Eq.~(\ref{stokes_law}) are given by
\begin{eqnarray}\label{U_i_0}
	\mathbf{U}_i^{(0)} ={} \frac{5-6\nu}{24\pi(1-\nu)\mu a}\mathbf{F}_i, \quad
	\boldsymbol{\Omega}_i^{(0)} ={} \frac{1}{8\pi\mu a^3}\mathbf{T}_i.
\end{eqnarray}
Corrections to the overall translation $\mathbf{U}_i$ and rotation $\boldsymbol{\Omega}_i$ of the $i$th sphere due to matrix-mediated interactions between the spheres are taken into account by adding them to $\mathbf{U}_i^{(0)}$ and $\boldsymbol{\Omega}_i^{(0)}$, respectively, in the form $\mathbf{U}_i=\mathbf{U}_i^{(0)}+\mathbf{U}_i^{(1)}+\dots$ and $\boldsymbol{\Omega}_i=\boldsymbol{\Omega}_i^{(0)}+\boldsymbol{\Omega}_i^{(1)}+\dots$.
This superposition works on the basis of the linearity of the underlying equations.
We can calculate these corrections from the displacement fields
\begin{eqnarray}
	\mathbf{u}_i^{(0)}(\mathbf{r}) &={}& \left(1+\frac{a^2}{6}\nabla^2\right)\mathbf{\underline{G}}(\mathbf{r}-\mathbf{r}_i)\cdot\mathbf{F}
	\notag\\
	&{}&-\frac{1}{2}(\mathbf{T}_i\times\nabla)\cdot\mathbf{\underline{G}}(\mathbf{r}-\mathbf{r}_i),\label{u_i_0}
\end{eqnarray}
and from the displacement fields
\begin{eqnarray}
	\mathbf{u}_k^{(n)}(\mathbf{r}) &={}& \sum\limits_{\scriptsize\begin{aligned}j\! &=\! 1 \\[-4pt] j\! &\ne \! k\end{aligned}}^{N}\bigg[-\mathbf{\underline{S}}_k^{(n)}(\mathbf{r}_{kj})\cdot\nabla\notag\\
	&{}&+\frac{1}{2}\left(\mathbf{\underline{\underline{M}}}_{k}^{(n)}(\mathbf{r}_{kj})-\frac{1}{3}\Tr{^{2\cdot3}\mathbf{\underline{\underline{M}}}_{k}^{(n)}(\mathbf{r}_{kj})}\mathbf{\hat{\underline{I}}}\right):\nabla\nabla\notag\\
	&{}&-\frac{1}{6}\mathbf{\underline{\underline{\underline{N}}}}_{k}^{(n)}(\mathbf{r}_{kj})\,\vdots\,\nabla\nabla\nabla\bigg]\cdot\mathbf{\underline{G}}(\mathbf{r}-\mathbf{r}_k),\label{eq_55}
\end{eqnarray}
with $n=1,2,3,\dots$, induced by the other spheres.
By the operator $\Tr{^{2\cdot3}}$ we imply contraction of the second and third index of the third-rank tensor $\mathbf{\underline{\underline{M}}}_k^{(n)}$.
We apply the Fax\'en relations, Eqs.~(\ref{faxen_F}) and (\ref{faxen_O}), to the displacement fields generated by the other spheres, so that the corrections to the translation $\mathbf{U}_i^{(0)}$ and rotation $\boldsymbol{\Omega}_i^{(0)}$ of a single sphere read
\begin{eqnarray}
	\mathbf{U}_i^{(n)} &={}& \sum\limits_{\scriptsize\begin{aligned}k\! &=\! 1 \\[-4pt] k\! &\ne \! i\end{aligned}}^{N}\left(1+\frac{a^2}{6}\nabla^2\right)\mathbf{u}_k^{(n-1)}(\mathbf{r})\bigg|_{\mathbf{r}=\mathbf{r}_i},\label{U_i}\\
	\boldsymbol{\Omega}_i^{(n)} &={}& \sum\limits_{\scriptsize\begin{aligned}k\! &=\! 1 \\[-4pt] k\! &\ne \! i\end{aligned}}^{N}\frac{1}{2}\nabla\times\mathbf{u}_k^{(n-1)}(\mathbf{r})|_{\mathbf{r}=\mathbf{r}_i},\label{omega_i}
\end{eqnarray}
$n\ge 1$.
The displacement field $\mathbf{u}_i^{(0)}(\mathbf{r})$ is in this context created directly by an external force $\mathbf{F}_i$ and/or an external torque $\mathbf{T}_i$.
In Eq.~(\ref{eq_55}), $\mathbf{r}_{kj}=\mathbf{r}_k-\mathbf{r}_j$,
\begin{eqnarray}
	\lefteqn{\mathbf{\underline{S}}_k^{(n)}(\mathbf{r}_{kj})}\notag\\ &={}& -\frac{4\pi(1-\nu)\mu a^3}{4-5\nu}\left(1+\frac{a^2}{10}\nabla^2\right)	 \bigg[
				\frac{1}{1-2\nu}\mathbf{\underline{\hat{I}}}\,\nabla\cdot\mathbf{u}_j^{(n-1)}(\mathbf{r})	\notag\\
		&{}&
				+\frac{5}{2}\Big(
				\nabla \mathbf{u}_j^{(n-1)}(\mathbf{r})
				+ \big(\nabla \mathbf{u}_j^{(n-1)}(\mathbf{r})\big)^\text{T}
				\Big)
			 \bigg]
			\bigg|_{\mathbf{r}=\mathbf{r}_k},\label{eq_56}
\end{eqnarray}
see Eq.~(\ref{S_jk}), and analogously for $\mathbf{\underline{\underline{M}}}_k^{(n)}(\mathbf{r}_{kj})$ and $\mathbf{\underline{\underline{\underline{N}}}}_k^{(n)}(\mathbf{r}_{kj})$, as obtained from Eqs.~(\ref{M_jkl_explicit}) and (\ref{N_jklm}), respectively.
By the summations, many-particle interactions are taken into account.
In this way, expressions for $\mathbf{U}_i$ and $\boldsymbol{\Omega}_i$ are obtained as functions of the inverse interparticle distances $r_{ik}^{-1}$, with $k=1,2,\dots,N$ and $k\neq i$.
We here stop at order $r_{ik}^{-6}$.
Thus, three-particle interactions are involved (four-particle interactions start to contribute only at order $r_{ik}^{-7}$).
That is, one particle generates a displacement field by deforming the surrounding matrix, initiated by a force or torque imposed on this particle.
A second particle resists the deformation implied by the resulting deformation field, which is expressed by the higher-order moments of its surface force density, see Eqs.~(\ref{def_S})--(\ref{def_N}).
Accordingly, a higher-order displacement field is generated, see Eq.~(\ref{eq_55}), to which a third particle (or again the first particle) is exposed.
For rigidly translated and/or rotated spherical inclusions, we thus stop at the level of $\mathbf{U}_i^{(2)}$ and $\boldsymbol{\Omega}_i^{(2)}$.
In the end, the effect of the coupled matrix-mediated interactions between the spheres can be written in the form of mathematical displaceability and rotateability matrices. 
Thus, the total translation and rotation of the $i$th particle to the given order are obtained as
\begin{eqnarray}
	\mathbf{U}_i &={} & \sum\limits_{j=1}^{N}\left[
		\mathbf{\underline{M}}_{ij}^\text{tt}\cdot\mathbf{F}_j + \mathbf{\underline{M}}_{ij}^\text{tr}\cdot\mathbf{T}_j
	 \right],\label{U_M}\\
	 \boldsymbol{\Omega}_i &={} & \sum\limits_{j=1}^{N}\left[
	 		\mathbf{\underline{M}}_{ij}^\text{rt}\cdot\mathbf{F}_j + \mathbf{\underline{M}}_{ij}^\text{rr}\cdot\mathbf{T}_j
	 	 \right],\label{OMEGA_M}
\end{eqnarray}
respectively, for $i=1,2,\dots,N$.

So far, the results up to (including) order $r_{ij}^{-4}$ had been calculated in Ref.~\onlinecite{puljiz2017forces}.
Our scope in the following subsections is to determine the contributions of the orders $r_{ij}^{-5}$ and $r_{ij}^{-6}$.
The complete expressions for the displaceability and rotateability matrices up to (including) order $r_{ij}^{-6}$ are for completeness listed as well.
\vspace{.5em}

\subsection{Displaceability matrices}
\label{A_displaceability}
First, we consider the situation, in which the $i$th particle is rigidly displaced by an external force $\mathbf{F}_i$.
This leads to the particle displacement $\mathbf{U}_i^{(0)}$, see Eq.~(\ref{U_i_0}), and induces the displacement field $\mathbf{u}_i^{(0)}(\mathbf{r})$ in the elastic medium, see Eq.~(\ref{u_i_0}).
The remaining contributions, to complete the order $r_{ik}^{-6}$ beyond the order $r_{ik}^{-4}$, see Ref.~\onlinecite{puljiz2017forces}, follow from the fields
\begin{eqnarray}
	\mathbf{u}_k^{(1)}(\mathbf{r}) &={}& \sum\limits_{\scriptsize\begin{aligned}j\! &=\! 1 \\[-4pt] j\! &\ne \! k\end{aligned}}^{N}\bigg[-\mathbf{\underline{S}}_k^{(1)}(\mathbf{r}_{kj})\cdot\nabla\notag\\
	&{}&
	+\frac{1}{2}\left(\mathbf{\underline{\underline{M}}}_{k}^{(1)}(\mathbf{r}_{kj})-\frac{1}{3}\Tr{^{2\cdot3}\mathbf{\underline{\underline{M}}}_{k}^{(1)}(\mathbf{r}_{kj})}\mathbf{\hat{\underline{I}}}\right):\nabla\nabla\notag\\
	&{}&-\frac{1}{6}\mathbf{\underline{\underline{\underline{N}}}}_{k}^{(1)}(\mathbf{r}_{kj})\,\vdots\,\nabla\nabla\nabla\bigg]\cdot\mathbf{\underline{G}}(\mathbf{r}-\mathbf{r}_k),
	\label{u_i_1}
\end{eqnarray}
see Eq.~(\ref{eq_55}).
\begin{widetext}
Here, the tensors are calculated via Eqs.~(\ref{S_jk}), (\ref{M_jkl_explicit}), and (\ref{N_jklm}) by inserting into Eq.~(\ref{eq_56}) and analogously into the corresponding expressions for $\mathbf{\underline{\underline{M}}}_k^{(1)}(\mathbf{r}_{kj})$ and $\mathbf{\underline{\underline{\underline{N}}}}_k^{(1)}(\mathbf{r}_{kj})$ the field $\mathbf{u}_j^{(0)}(\mathbf{r})$ obtained via Eq.~(\ref{u_i_0}).
For $\mathbf{\underline{S}}_k^{(1)}(\mathbf{r}_{kj})$, we find
\begin{eqnarray}
	\mathbf{\underline{S}}_k^{(1)}(\mathbf{r}_{kj})
	&={} & -\frac{1}{4(4-5\nu)}\frac{a^3}{r_{kj}^2}\Bigg[
		3\mathbf{\hat{\underline{I}}}\,\mathbf{\hat{r}}_{kj}\cdot\mathbf{F}_j -5(1-2\nu)(\mathbf{F}_j \mathbf{\hat{r}}_{kj} + \mathbf{\hat{r}}_{kj} \mathbf{F}_j)
		-15\mathbf{\hat{r}}_{kj}\mathbf{\hat{r}}_{kj}\mathbf{\hat{r}}_{kj}\cdot\mathbf{F}_j
		+8\left(\frac{a}{r_{kj}}\right)^2\Big(5 \mathbf{\hat{r}}_{kj} \mathbf{\hat{r}}_{kj} \mathbf{\hat{r}}_{kj}\cdot\mathbf{F}_j\notag\\
		&{}&
		-(\mathbf{F}_j\mathbf{\hat{r}}_{kj}+\mathbf{\hat{r}}_{kj}\mathbf{F}_j+\mathbf{\hat{\underline{I}}}\,\mathbf{\hat{r}}_{kj}\cdot\mathbf{F}_j)\Big)
	\Bigg].\label{stresslet_displ}
\end{eqnarray}
Therefore, the sixth-order contribution to $\mathbf{U}_i^{(2)}$ arising from the stresslet generated on particle $k$ due to its resistance to deformation in a displacement field generated by a force $\mathbf{F}_j$ acting on particle $j$ is obtained from
\begin{equation}
 -\left(1+\frac{a^2}{6}\nabla^2\right)\Big(\mathbf{\underline{S}}_k^{(1)}(\mathbf{r}_{kj})\cdot\nabla\Big)\cdot\mathbf{\underline{G}}(\mathbf{r}-\mathbf{r}_k)\bigg|_{\mathbf{r}=\mathbf{r}_i}
\end{equation}
as
\begin{eqnarray}
 &{}& M_0^\text{t}\frac{3}{8(4-5\nu)(5-6\nu)}\Bigg\{
 5\left(\frac{a}{r_{ik}}\right)^4\left(\frac{a}{r_{jk}}\right)^2
 \Big[
  2(1-2\nu)\Big(
  	5 \mathbf{\hat{r}}_{ik}\mathbf{\hat{r}}_{ik}(\mathbf{\hat{r}}_{ik}\cdot\mathbf{\hat{r}}_{jk})
  	 \notag\\
  	 &{}&-\big(
  		\mathbf{\hat{\underline{I}}}\,(\mathbf{\hat{r}}_{ik}\cdot\mathbf{\hat{r}}_{jk})
  	+\mathbf{\hat{r}}_{ik}\mathbf{\hat{r}}_{jk}+\mathbf{\hat{r}}_{jk}\mathbf{\hat{r}}_{ik}\big)
  \Big)+15\mathbf{\hat{r}}_{ik}\mathbf{\hat{r}}_{jk}(\mathbf{\hat{r}}_{ik}\cdot\mathbf{\hat{r}}_{jk})^2-6\mathbf{\hat{r}}_{jk}\mathbf{\hat{r}}_{jk}(\mathbf{\hat{r}}_{ik}\cdot\mathbf{\hat{r}}_{jk})
  -3\mathbf{\hat{r}}_{ik}\mathbf{\hat{r}}_{jk}
  \Big]\notag\\
  &{}&
 +8\left(\frac{a}{r_{ik}}\right)^2\left(\frac{a}{r_{jk}}\right)^4\Big[
 2(1-2\nu)\Big(
 	5 \mathbf{\hat{r}}_{jk}\mathbf{\hat{r}}_{jk}(\mathbf{\hat{r}}_{ik}\cdot\mathbf{\hat{r}}_{jk})
 	 -\big(
 		\mathbf{\hat{\underline{I}}}\,(\mathbf{\hat{r}}_{ik}\cdot\mathbf{\hat{r}}_{jk})
 	\notag\\
 	 	 &{}&
 	+\mathbf{\hat{r}}_{ik}\mathbf{\hat{r}}_{jk}+\mathbf{\hat{r}}_{jk}\mathbf{\hat{r}}_{ik}
 	\big)
 \Big)
 +15\mathbf{\hat{r}}_{ik}\mathbf{\hat{r}}_{jk}(\mathbf{\hat{r}}_{ik}\cdot\mathbf{\hat{r}}_{jk})^2
 -6\mathbf{\hat{r}}_{ik}\mathbf{\hat{r}}_{ik}(\mathbf{\hat{r}}_{ik}\cdot\mathbf{\hat{r}}_{jk})
 -3\mathbf{\hat{r}}_{ik}\mathbf{\hat{r}}_{jk}
 \Big]\Bigg\}
 \cdot\mathbf{F}_j,
\end{eqnarray}
with $M_0^\text{t}=(5-6\nu)/24\pi(1-\nu)\mu a$.
(There is no fifth-order contribution involving $\mathbf{\underline{S}}_{k}^{(1)}$.)

Next, the $\mathbf{\underline{\underline{M}}}_{k}^{(1)}(\mathbf{r}_{kj})$-term is addressed.
We find from
\begin{equation}
	M_{jkl}^{k(1)}(\mathbf{r}^{kj}) - \frac{a^2}{3}F_{j}^j\delta_{kl}
\end{equation}
in Eq.~(\ref{M_jkl_explicit}) to third order the contribution
\begin{eqnarray}
	&{}&\frac{1}{6(11-14\nu)}\frac{a^5}{(r^{kj})^3}
	\Bigg\{
		-105 \hat{r}_{j}^{kj}\hat{r}_{k}^{kj}\hat{r}_{l}^{kj}\hat{r}_{m}^{kj}+(17-28\nu)\delta_{jm}\left(\delta_{kl}
		-3 \hat{r}_{k}^{kj}\hat{r}_{l}^{kj}
		\right)\notag\\
		&{}&
		-5\left(
		\delta_{jk}\delta_{lm}+\delta_{jl}\delta_{km}
		-3\delta_{lm} \hat{r}_{j}^{kj}\hat{r}_{k}^{kj}
		-3\delta_{km} \hat{r}_{j}^{kj}\hat{r}_{l}^{kj}
		\right)
		+21\left(
			\delta_{jk} \hat{r}_{l}^{kj}\hat{r}_{m}^{kj}
			+\delta_{jl} \hat{r}_{k}^{kj}\hat{r}_{m}^{kj}
			+\delta_{kl} \hat{r}_{j}^{kj}\hat{r}_{m}^{kj}
		\right)
		\notag\\
		&{}&
		+2 \Big(
			\delta_{jk}(\delta_{lm}-3\hat{r}_{l}^{kj}\hat{r}_{m}^{kj})
		+\delta_{jl}(\delta_{km}-3\hat{r}_{k}^{kj}\hat{r}_{m}^{kj})
	+\delta_{kl}(\delta_{jm}-3\hat{r}_{j}^{kj}\hat{r}_{m}^{kj})
		\Big)
	\Bigg\}F_{m}^{j}
	.
\end{eqnarray}
Here, for better readability, we have shifted the particle indices to the superscript, which are not summed over, and the coordinate indices to the subscript.
Thus, 
we obtain to sixth order from
\begin{equation}
	\frac{1}{2}\left[\left(\mathbf{\underline{\underline{M}}}_{k}^{(1)}(\mathbf{r}_{kj})-\frac{1}{3}\Tr{^{2\cdot3}\mathbf{\underline{\underline{M}}}_{k}^{(1)}}(\mathbf{r}_{kj})\mathbf{\hat{\underline{I}}}\right):\nabla\nabla\right]\cdot\mathbf{\underline{G}}(\mathbf{r}-\mathbf{r}_k)\bigg|_{\mathbf{r}=\mathbf{r}_i}
\end{equation}
the expression
\begin{eqnarray}
	&{} & M_0^\text{t} \frac{1}{8(5-6\nu)(11-14\nu)} \left(\frac{a}{r_{ik}}\right)^3\left(\frac{a}{r_{jk}}\right)^3
	\Bigg\{
	-1575 \mathbf{\hat{r}}_{ik}\mathbf{\hat{r}}_{jk} (\mathbf{\hat{r}}_{ik}\cdot\mathbf{\hat{r}}_{jk})^3
	\notag\\
	&{}&
	-315(1-4\nu)\left[
		\mathbf{\hat{r}}_{ik}\mathbf{\hat{r}}_{ik}(\mathbf{\hat{r}}_{ik}\cdot\mathbf{\hat{r}}_{jk})^2
		+\mathbf{\hat{r}}_{jk}\mathbf{\hat{r}}_{jk}(\mathbf{\hat{r}}_{ik}\cdot\mathbf{\hat{r}}_{jk})^2
	\right]\notag\\
	&{}&
	+27(43-32\nu) \mathbf{\hat{r}}_{ik}\mathbf{\hat{r}}_{jk}(\mathbf{\hat{r}}_{ik}\cdot\mathbf{\hat{r}}_{jk})+54(9-16\nu)\mathbf{\hat{r}}_{jk}\mathbf{\hat{r}}_{ik}(\mathbf{\hat{r}}_{ik}\cdot\mathbf{\hat{r}}_{jk})
	-3(29+4\nu)\left[
		\mathbf{\hat{r}}_{ik}\mathbf{\hat{r}}_{ik}+\mathbf{\hat{r}}_{jk}\mathbf{\hat{r}}_{jk}
	\right]
	\notag\\
	&{}&
	-9(61-152\nu+112\nu^2) \mathbf{\hat{\underline{I}}} \,(\mathbf{\hat{r}}_{ik}\cdot\mathbf{\hat{r}}_{jk})^2
	+(233-536\nu+336\nu^2)\mathbf{\hat{\underline{I}}}
	\Bigg\}\cdot\mathbf{F}_j
	.
\end{eqnarray}

As for $\mathbf{\underline{\underline{\underline{N}}}}_k^{(1)}(\mathbf{r}_{kj})$, we find 
via Eq.~(\ref{N_identity}) and the stresslet in Eq.~(\ref{stresslet_displ}) that from $-N_{jklm}^{k(1)}(\mathbf{r}^{kj})/6$ only
\begin{eqnarray}
		&{}&-\frac{1}{8(4-5\nu)}\frac{a^5}{(r^{kj})^2}\Big\{
		3 \hat{r}_{j}^{kj}\hat{r}_{k}^{kj}\mathbf{\hat{r}}^{kj}\cdot\mathbf{F}^j
		+(1-2\nu)[ F_{j}^j\hat{r}_{k}^{kj} + \hat{r}_{j}^{kj} F_{k}^{j} ]
	\Big\}
\end{eqnarray}
adds to our considered order.
This leads to a sixth-order contribution from
\begin{equation}
	-\frac{1}{6}\left[\mathbf{\underline{\underline{\underline{N}}}}_{k}^{(1)}(\mathbf{r}_{kj})\,\vdots\,\nabla\nabla\nabla\right]\cdot \mathbf{\underline{G}}(\mathbf{r}-\mathbf{r}_k)\bigg|_{\mathbf{r}=\mathbf{r}_i}
\end{equation}
of the form
\begin{eqnarray}
	&{}& M_0^\text{t}\frac{9}{8(4-5\nu)(5-6\nu)}\left(\frac{a}{r_{ik}}\right)^4\left(\frac{a}{r_{jk}}\right)^2
	\Bigg\{
	15\mathbf{\hat{r}}_{ik}\mathbf{\hat{r}}_{jk}(\mathbf{\hat{r}}_{ik}\cdot\mathbf{\hat{r}}_{jk})^2
	-6 \mathbf{\hat{r}}_{jk}\mathbf{\hat{r}}_{jk}(\mathbf{\hat{r}}_{ik}\cdot\mathbf{\hat{r}}_{jk})-(5-4\nu)\mathbf{\hat{r}}_{ik}\mathbf{\hat{r}}_{jk}	\notag\\
	&{}&-2(1-2\nu)\Big[
		\left(\mathbf{\hat{\underline{I}}}
		-5\mathbf{\hat{r}}_{ik}\mathbf{\hat{r}}_{ik}\right)(\mathbf{\hat{r}}_{ik}\cdot\mathbf{\hat{r}}_{jk})
		+\mathbf{\hat{r}}_{jk}\mathbf{\hat{r}}_{ik}
	\Big]
	\Bigg\}\cdot\mathbf{F}_j.
\end{eqnarray}

The additional rotation $\boldsymbol{\Omega}_i^{(2)}$ up to (including) sixth order is obtained solely from the stresslet term and, via Eqs.~(\ref{omega_i}), (\ref{u_i_1}), and (\ref{stresslet_displ}), reads
\begin{eqnarray}
	\boldsymbol{\Omega}_i^{(2)} &={}& M_0^\text{r} \frac{15}{4(4-5\nu)}\sum\limits_{\scriptsize\begin{aligned}j,k\! &=\! 1 \\[-4pt] k\! &\ne \! i,\!j\end{aligned}}^{N}
	\frac{a^3}{r_{ik}^3 r_{jk}^2}\Big\{
	(1-2\nu)\Big[
		 \mathbf{\hat{r}}_{ik}\times\mathbf{\hat{\underline{I}}}\,(\mathbf{\hat{r}}_{ik}\cdot\mathbf{\hat{r}}_{jk})
		+ (\mathbf{\hat{r}}_{ik}\times\mathbf{\hat{r}}_{jk})\mathbf{\hat{r}}_{ik}
	\Big] + 3(\mathbf{\hat{r}}_{ik}\times\mathbf{\hat{r}}_{jk})\mathbf{\hat{r}}_{jk} (\mathbf{\hat{r}}_{ik}\cdot\mathbf{\hat{r}}_{jk})
	\Big\}\cdot\mathbf{F}_j,\qquad
\end{eqnarray}
with $M_0^\text{r}=1/8\pi\mu$ (there is only this additional fifth-order contribution up to the considered level).

In total, together with our previously derived expressions to fourth order \cite{puljiz2017forces}, the sixth-order displaceability and rotateability matrices for translation--translation and rotation--translation couplings, respectively, read
\begin{eqnarray}
	\mathbf{\underline{M}}_{i=j}^\text{tt} &={}& M_0^\text{t}\,\mathbf{\underline{\hat{I}}}+\mathbf{\underline{M}}_{ij}^{\text{tt}(4,6)},	\\
	\mathbf{\underline{M}}_{i\not=j}^\text{tt} &={} &  M_0^\text{t}\frac{3}{2(5-6\nu)}\frac{a}{r_{ij}}  \Bigg\{ \Bigg( 4(1-\nu)-\frac{4}{3} \bigg(\frac{a}{r_{ij}}\bigg)^2 \Bigg) \mathbf{\hat{r}}_{ij}\mathbf{\hat{r}}_{ij}   + \Bigg( 3-4\nu+\frac{2}{3} \bigg(\frac{a}{r_{ij}} \bigg)^2 \Bigg)(\mathbf{\underline{\hat{I}}}-\mathbf{\hat{r}}_{ij}\mathbf{\hat{r}}_{ij}) \Bigg\}+\mathbf{\underline{M}}_{ij}^{\text{tt}(4,6)},\\
	\mathbf{\underline{M}}_{ij}^{\text{tt}(4,6)} &={} & M_0^\text{t}
	\sum\limits_{\scriptsize\begin{aligned}k\! &=\! 1 \\[-4pt] k\! &\ne \! i,\!j\end{aligned}}^{N}\Bigg\{
	\frac{3}{8(4-5\nu)(5-6\nu)}\Bigg(
	\left(\frac{a}{r_{ik}}\right)^2\left(\frac{a}{r_{jk}}\right)^2\Big[
		-10(1-2\nu)\Big(
			(1-2\nu)\big( \mathbf{\hat{\underline{I}}}\,(\mathbf{\hat{r}}_{ik}\cdot\mathbf{\hat{r}}_{jk})+\mathbf{\hat{r}}_{jk}\mathbf{\hat{r}}_{ik}\big)
			\notag\\
		&{}&
		+3(\mathbf{\hat{r}}_{ik}\cdot\mathbf{\hat{r}}_{jk})(\mathbf{\hat{r}}_{ik}\mathbf{\hat{r}}_{ik}+\mathbf{\hat{r}}_{jk}\mathbf{\hat{r}}_{jk}) - \mathbf{\hat{r}}_{ik}\mathbf{\hat{r}}_{jk}
		\Big)
		+3\big(
			7-4\nu-15(\mathbf{\hat{r}}_{ik}\cdot\mathbf{\hat{r}}_{jk})^2
		\big)\mathbf{\hat{r}}_{ik}\mathbf{\hat{r}}_{jk}
	\Big]
	\notag\\
	&{}&+
	 2
	 \left(\frac{a}{r_{ik}}\right)^4\left(\frac{a}{r_{jk}}\right)^2
	  \Big[
	  (1-2\nu)\Big(
	   	40 \mathbf{\hat{r}}_{ik}\mathbf{\hat{r}}_{ik}(\mathbf{\hat{r}}_{ik}\cdot\mathbf{\hat{r}}_{jk})-\big(
	   		8\mathbf{\hat{\underline{I}}}\,(\mathbf{\hat{r}}_{ik}\cdot\mathbf{\hat{r}}_{jk})
	   	+5\mathbf{\hat{r}}_{ik}\mathbf{\hat{r}}_{jk}+8\mathbf{\hat{r}}_{jk}\mathbf{\hat{r}}_{ik}
	   		   	\big)
	   \Big)
	   	   	 \notag\\
	   	   	 &{}&
	   	   	 +3\Big(20\mathbf{\hat{r}}_{ik}\mathbf{\hat{r}}_{jk}(\mathbf{\hat{r}}_{ik}\cdot\mathbf{\hat{r}}_{jk})^2
	   	   	 -8\mathbf{\hat{r}}_{jk}\mathbf{\hat{r}}_{jk}(\mathbf{\hat{r}}_{ik}\cdot\mathbf{\hat{r}}_{jk})
	   -(5-2\nu)\mathbf{\hat{r}}_{ik}\mathbf{\hat{r}}_{jk}
	   \Big)
	   \Big]\notag\\
	   &{}&
	  +8\left(\frac{a}{r_{ik}}\right)^2\left(\frac{a}{r_{jk}}\right)^4\Big[
	  2(1-2\nu)\Big(
	  	5 \mathbf{\hat{r}}_{jk}\mathbf{\hat{r}}_{jk}(\mathbf{\hat{r}}_{ik}\cdot\mathbf{\hat{r}}_{jk})
	  	 -\big(
	  		\mathbf{\hat{\underline{I}}}\,(\mathbf{\hat{r}}_{ik}\cdot\mathbf{\hat{r}}_{jk})
	  	\notag\\
	  	 	 &{}&
	  	+\mathbf{\hat{r}}_{ik}\mathbf{\hat{r}}_{jk}+\mathbf{\hat{r}}_{jk}\mathbf{\hat{r}}_{ik}\big)
	  \Big)
	  +15\mathbf{\hat{r}}_{ik}\mathbf{\hat{r}}_{jk}(\mathbf{\hat{r}}_{ik}\cdot\mathbf{\hat{r}}_{jk})^2
	  -6\mathbf{\hat{r}}_{ik}\mathbf{\hat{r}}_{ik}(\mathbf{\hat{r}}_{ik}\cdot\mathbf{\hat{r}}_{jk})
	  -3\mathbf{\hat{r}}_{ik}\mathbf{\hat{r}}_{jk}
	  \Big]\Bigg)\notag\\
	 &{}&+
	 \frac{1}{8(5-6\nu)(11-14\nu)} \left(\frac{a}{r_{ik}}\right)^3\left(\frac{a}{r_{jk}}\right)^3
	 	\Big(
	 	-1575 \mathbf{\hat{r}}_{ik}\mathbf{\hat{r}}_{jk} (\mathbf{\hat{r}}_{ik}\cdot\mathbf{\hat{r}}_{jk})^3
	 		 	\notag\\
	 		 	&{}&
	 	-315(1-4\nu)\left[
	 		\mathbf{\hat{r}}_{ik}\mathbf{\hat{r}}_{ik}(\mathbf{\hat{r}}_{ik}\cdot\mathbf{\hat{r}}_{jk})^2
	 		+\mathbf{\hat{r}}_{jk}\mathbf{\hat{r}}_{jk}(\mathbf{\hat{r}}_{ik}\cdot\mathbf{\hat{r}}_{jk})^2
	 	\right]\notag\\
	 	&{}&
	 	+27(43-32\nu) \mathbf{\hat{r}}_{ik}\mathbf{\hat{r}}_{jk}(\mathbf{\hat{r}}_{ik}\cdot\mathbf{\hat{r}}_{jk})+54(9-16\nu)\mathbf{\hat{r}}_{jk}\mathbf{\hat{r}}_{ik}(\mathbf{\hat{r}}_{ik}\cdot\mathbf{\hat{r}}_{jk})
	 	-3(29+4\nu)\left[
	 		\mathbf{\hat{r}}_{ik}\mathbf{\hat{r}}_{ik}+\mathbf{\hat{r}}_{jk}\mathbf{\hat{r}}_{jk}
	 	\right]
	 	\notag\\
	 	&{}&
	 	-9(61-152\nu+112\nu^2) \mathbf{\hat{\underline{I}}} \,(\mathbf{\hat{r}}_{ik}\cdot\mathbf{\hat{r}}_{jk})^2
	 	+(233-536\nu+336\nu^2)\mathbf{\hat{\underline{I}}}
	 	\Big) 
		\Bigg\},\label{M_ij_tt_46}
\end{eqnarray}
and
\begin{eqnarray}
	\mathbf{\underline{M}}_{i=j}^\text{rt} &={}&\mathbf{\underline{M}}_{ij}^{\text{rt}(5)},\label{M_rt_0}\\
	\mathbf{\underline{M}}_{i\not=j}^\text{rt} &={}& -M_0^\text{r} \frac{\mathbf{\hat{r}}_{ij}}{r_{ij}^2}\times\mathbf{\hat{\underline{I}}} + \mathbf{\underline{M}}_{ij}^{\text{rt}(5)},\label{M_rt_2}\\
	\mathbf{\underline{M}}_{ij}^{\text{rt}(5)} &={} & M_0^\text{r}\frac{15}{4(4-5\nu)}\sum\limits_{\scriptsize\begin{aligned}k\! &=\! 1 \\[-4pt] k\! &\ne \! i,\!j\end{aligned}}^{N}
		\frac{a^3}{r_{ik}^3 r_{jk}^2}\Big\{
		(1-2\nu)\Big[
			 \mathbf{\hat{r}}_{ik}\times\mathbf{\hat{\underline{I}}}\,(\mathbf{\hat{r}}_{ik}\cdot\mathbf{\hat{r}}_{jk})
			+ (\mathbf{\hat{r}}_{ik}\times\mathbf{\hat{r}}_{jk})\mathbf{\hat{r}}_{ik}
		\Big] + 3(\mathbf{\hat{r}}_{ik}\times\mathbf{\hat{r}}_{jk})\mathbf{\hat{r}}_{jk} (\mathbf{\hat{r}}_{ik}\cdot\mathbf{\hat{r}}_{jk})
		\Big\}.\qquad\label{M_rt_5}
\end{eqnarray}

\subsection{Rotateability matrices}
\label{B_rotateability}
Likewise, an external torque $\mathbf{T}_i$ acting on the $i$th particle to zeroth order leads to a displacement field $\mathbf{u}_i^{(0)}(\mathbf{r})$ as given by Eq.~(\ref{u_i_0}) and the particle rotation $\boldsymbol{\Omega}_i^{(0)}$ as specified by Eq.~(\ref{U_i_0}).
Overall, up to (including) order $r_{ij}^{-6}$, only third-order parts of the stresslets $\mathbf{\underline{S}}_{k}^{(1)}(\mathbf{r}_{kj})$ contribute to $\mathbf{U}_i^{(2)}$ and $\boldsymbol{\Omega}_i^{(2)}$, which here according to Eqs.~(\ref{u_i_0}) and (\ref{eq_56}) are of the form
\begin{equation}
	\frac{15(1-\nu)}{4(4-5\nu)}\left(\frac{a}{r_{kj}}\right)^3\Big\{
		\mathbf{\hat{r}}_{kj} \mathbf{T}_j\times\mathbf{\hat{r}}_{kj}
		+ \mathbf{T}_j\times\mathbf{\hat{r}}_{kj}\mathbf{\hat{r}}_{kj}
	\Big\}.
\end{equation}
They yield the additional contributions to the translation and rotation, respectively, of
\begin{eqnarray}
	\mathbf{U}_i^{(2)} &={} & -M_0^\text{r}\frac{15}{4(4-5\nu)}\sum\limits_{\scriptsize\begin{aligned}j,k\! &=\! 1 \\[-4pt] k\! &\ne \! i,\!j\end{aligned}}^{N}\frac{a^3}{r_{ik}^2r_{jk}^3}\Big\{
	(1-2\nu)\Big[
		\mathbf{\hat{r}}_{jk}\times\mathbf{\hat{\underline{I}}}\,(\mathbf{\hat{r}}_{ik}\cdot\mathbf{\hat{r}}_{jk})
		+
		\mathbf{\hat{r}}_{jk}(\mathbf{\hat{r}}_{ik}\times\mathbf{\hat{r}}_{jk})
	\Big]
	+3\mathbf{\hat{r}}_{ik} 
	(\mathbf{\hat{r}}_{ik}\times\mathbf{\hat{r}}_{jk})
			(\mathbf{\hat{r}}_{ik}\cdot\mathbf{\hat{r}}_{jk})
	\Big\}\cdot\mathbf{T}_j\qquad
\end{eqnarray}
according to Eqs.~(\ref{U_i}) and (\ref{u_i_1}), and of
\begin{eqnarray}
	\boldsymbol{\Omega}_i^{(2)} &={}& M_0^\text{r} \frac{45(1-\nu)}{4(4-5\nu)}\sum\limits_{\scriptsize\begin{aligned}j,k\! &=\! 1 \\[-4pt] k\! &\ne \! i,\!j\end{aligned}}^{N}\frac{a^3}{r_{ik}^3r_{jk}^3}\Big\{
		(\mathbf{\hat{r}}_{ik}\times\mathbf{\hat{r}}_{jk}) (\mathbf{\hat{r}}_{ik}\times\mathbf{\hat{r}}_{jk})
		+\mathbf{\hat{r}}_{jk}\mathbf{\hat{r}}_{ik}(\mathbf{\hat{r}}_{ik}\cdot\mathbf{\hat{r}}_{jk})
		-\mathbf{\underline{\hat{I}}}\, (\mathbf{\hat{r}}_{ik}\cdot\mathbf{\hat{r}}_{jk})^2
		\Big\}\cdot\mathbf{T}_j
\end{eqnarray}
according to Eqs.~(\ref{omega_i}) and (\ref{u_i_1}) (there is no sixth-order contribution to the translation resulting from an imposed torque).
We stress that the sixth-order contribution to the rotation is nonzero also for incompressible systems, i.e., for $\nu\rightarrow1/2$.

In summary, the translation--rotation and rotation--rotation coupling matrices, together with our previously-derived expressions to fourth order \cite{puljiz2017forces}, are given by
\begin{eqnarray}
	\mathbf{\underline{M}}_{i=j}^\text{tr} &={}& \mathbf{\underline{M}}_{ij}^{\text{tr}(5)},\\
	\mathbf{\underline{M}}_{i\not=j}^\text{tr} &={}& -M_0^\text{r}\frac{\mathbf{\hat{r}}_{ij}}{r_{ij}^2}\times\mathbf{\underline{\hat{I}}}+\mathbf{\underline{M}}_{ij}^{\text{tr}(5)},\\
	\mathbf{\underline{M}}_{ij}^{\text{tr}(5)} &={} & -M_0^\text{r}\frac{15}{4(4-5\nu)}\sum\limits_{\scriptsize\begin{aligned}k\! &=\! 1 \\[-4pt] k\! &\ne \! i,\!j\end{aligned}}^{N}\frac{a^3}{r_{ik}^2r_{jk}^3}\Big\{
		(1-2\nu)\Big[
			\mathbf{\hat{r}}_{jk}\times\mathbf{\hat{\underline{I}}}\,(\mathbf{\hat{r}}_{ik}\cdot\mathbf{\hat{r}}_{jk})
			+
			\mathbf{\hat{r}}_{jk}(\mathbf{\hat{r}}_{ik}\times\mathbf{\hat{r}}_{jk})
		\Big]
		+3\mathbf{\hat{r}}_{ik} 
		(\mathbf{\hat{r}}_{ik}\times\mathbf{\hat{r}}_{jk})
				(\mathbf{\hat{r}}_{ik}\cdot\mathbf{\hat{r}}_{jk})
		\Big\},\qquad
\end{eqnarray}
and
\begin{eqnarray}
		\mathbf{\underline{M}}_{i=j}^\text{rr} &={}& M_0^\text{r}\frac{1}{a^3}\mathbf{\underline{\hat{I}}} + \mathbf{\underline{M}}_{ij}^{\text{rr}(6)},\label{M_i=j_rr}\\
		\mathbf{\underline{M}}_{i\not=j}^\text{rr} &={}& M_0^\text{r}\frac{1}{2r_{ij}^3}\big[3\mathbf{\hat{r}}_{ij}\mathbf{\hat{r}}_{ij}-\mathbf{\underline{\hat{I}}}\big]+\mathbf{\underline{M}}_{ij}^{\text{rr}(6)},
				\label{M_ineqj_rr}
		\\
	\mathbf{\underline{M}}_{ij}^{\text{rr}(6)} &={} &  M_0^\text{r} \frac{45(1-\nu)}{4(4-5\nu)}\sum\limits_{\scriptsize\begin{aligned}k\! &=\! 1 \\[-4pt] k\! &\ne \! i,\!j\end{aligned}}^{N}\frac{a^3}{r_{ik}^3r_{jk}^3}\Big\{
			(\mathbf{\hat{r}}_{ik}\times\mathbf{\hat{r}}_{jk}) (\mathbf{\hat{r}}_{ik}\times\mathbf{\hat{r}}_{jk})
			+\mathbf{\hat{r}}_{jk}\mathbf{\hat{r}}_{ik}(\mathbf{\hat{r}}_{ik}\cdot\mathbf{\hat{r}}_{jk})
			-\mathbf{\underline{\hat{I}}}\, (\mathbf{\hat{r}}_{ik}\cdot\mathbf{\hat{r}}_{jk})^2
			\Big\},
			\label{M_ij_rr_6}
\end{eqnarray}
respectively.

For all derived displaceability and rotateability matrices, the expressions of corresponding low-Reynolds-number hydrodynamic mobility matrices for incompressible fluid flows are obtained by setting $\nu=1/2$ in the limit of an incompressible environment and replacing $\mu$ by the hydrodynamic viscosity $\eta$.
Previously, for such incompressible fluid surroundings, the expressions for the hydrodynamic mobility matrices had been derived in a different way \cite{mazur1982many}.
These expressions are recovered from ours for $\nu=1/2$.
\end{widetext}

\subsection{Examples: Magnetic particles}\label{section_examples}

We now present examples for two- and three-particle systems. Specifically, the difference between particle interactions through an elastic compressible or an incompressible environment is illustrated, together with the effect of the higher-order terms.
For simplicity, we assume magnetic dipole--dipole interactions between $N$ identical particles embedded in an elastic medium, reminiscent of the situation in ferrogels and magnetorheological elastomers \cite{jarkova2003hydrodynamics,attaran2017modeling,weeber2018studying,metsch2018two}. 
Corresponding magnetic dipole forces are given by \cite{jackson1962classical}
\begin{equation}
	\mathbf{F}_i ={} -\frac{3\mu_0 m^2}{4\pi}\sum\limits_{\substack{j=1\\j\neq i}}^{N} \frac{5\mathbf{\hat{r}}_{ij}(\mathbf{\hat{m}}\cdot\mathbf{\hat{r}}_{ij})^2-\mathbf{\hat{r}}_{ij}-2\mathbf{\hat{m}}(\mathbf{\hat{m}}\cdot\mathbf{\hat{r}}_{ij})}{r_{ij}^4},
	\label{dip-dip_f}
\end{equation}
$i=1,\dots,N$.
Here, $\mu_0$ is the magnetic vacuum permeability and $\mathbf{m}$ represents the magnetic dipole moment that we assume to be identical for all particles ($m=|\mathbf{m}|$, $\mathbf{\hat{m}}=\mathbf{m}/m$). 
This can be achieved, for instance, by a strong homogeneous saturating external magnetic field $\mathbf{B}$ magnetizing the spherical particles \cite{puljiz2016forces},
with the magnetic moments oriented along $\mathbf{B}$.
If particle displacements occur, Eq.~(\ref{dip-dip_f}) should be evaluated with the modified particle distances considered \cite{puljiz2016forces}.
An iterative numerical loop is employed for this purpose to calculate the final displacements \cite{puljiz2016forces}, which in the regime of linear elasticity converges quickly.
In the following, we set $\mu_0 m^2/\mu a^6 = 5\times 10^2$. 
Moreover, all initial interparticle center-to-center distances are set to $3a$.

Assuming uniaxial magnetic anisotropy of the particles, magnetic torques are calculated from the idealized Stoner--Wohlfarth model \cite{stoner1948mechanism,puljiz2017forces,roeder2015magnetic},
\begin{equation}\label{stoner-wohlfarth}
	\mathbf{T}_i ={} \frac{8\pi a^3}{3}K(\mathbf{\hat{n}}_i\cdot\mathbf{\hat{B}}) \mathbf{\hat{n}}_i\times \mathbf{\hat{B}}.
\end{equation}
In this expression, $\mathbf{\hat{B}}=\mathbf{B}/|\mathbf{B}|$ and $\mathbf{\hat{n}}_i$ represents the nonpolar axis of magnetic anisotropy of the $i$th particle, rigidly anchored to the particle frame.
We set the anisotropy parameter $K/\mu=1/3$.

\begin{figure}
\centerline{\includegraphics[width=\columnwidth]{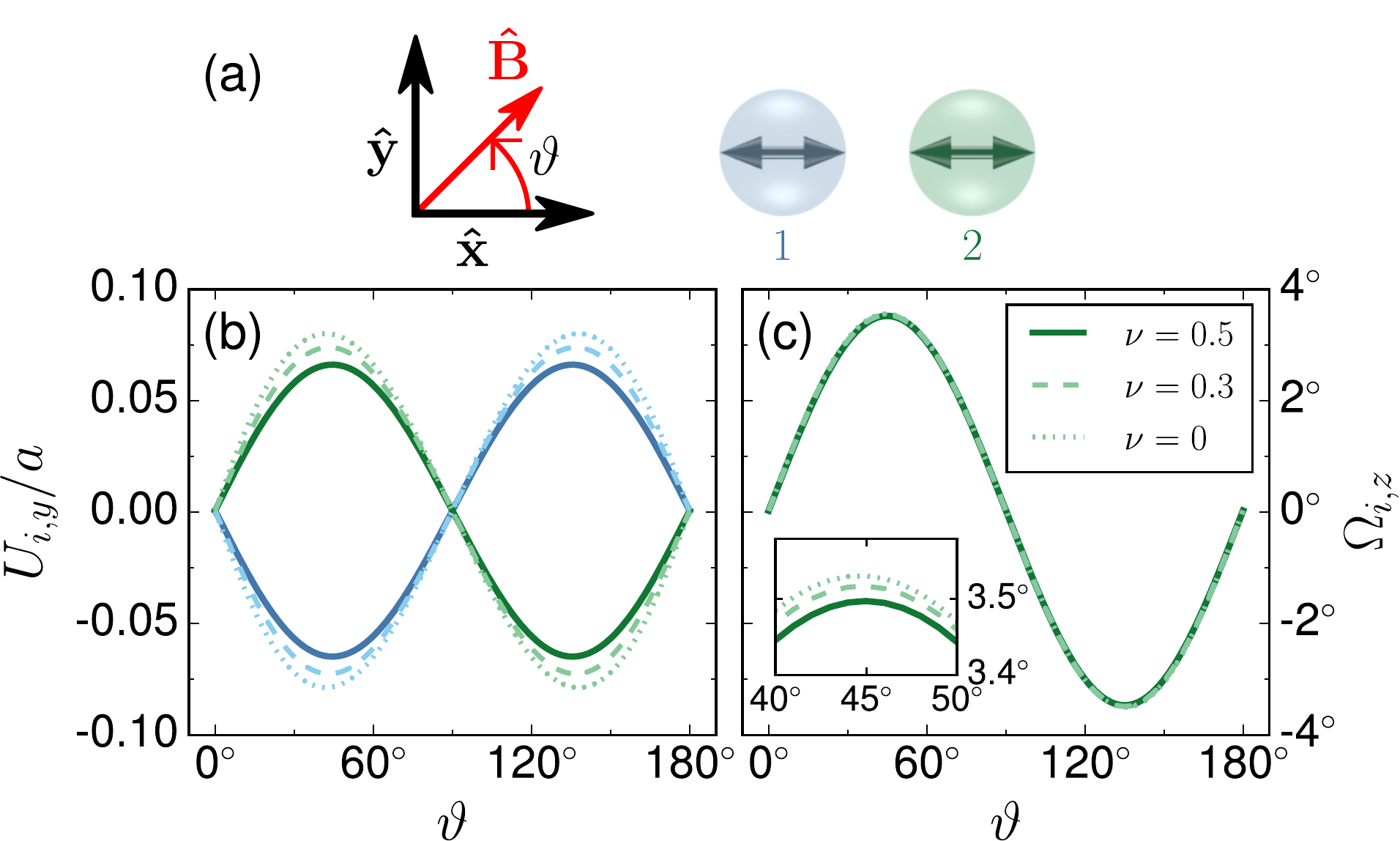}}
\caption{Two-particle system ($i=1,2$) in the $xy$-plane, in which both magnetic forces and torques are induced by a strong external magnetic field of direction $\mathbf{\hat{B}}$ rotated counterclockwise in the $xy$-plane by an angle $\vartheta$ relatively to the $\mathbf{\hat{x}}$-axis, see Eqs.~(\ref{dip-dip_f}) and (\ref{stoner-wohlfarth}). 
(a) Sketch of the system configuration.
The double-headed arrows in the spheres indicate the orientation of the magnetic anisotropy axes $\mathbf{\hat{n}}_i$ for $\mathbf{B}=\mathbf{0}$.
(b) $y$-component of the particle displacements $\mathbf{U}_i$.
The solid, dashed, and dotted lines represent $U_{i,y}$ for Poisson ratios $\nu=0.5$, $0.3$, and $0$, respectively.
(c) $z$-component of the particle rotations $\boldsymbol{\Omega}_i$.
Analogously, the solid, dashed, and dotted lines show $\Omega_{i,z}$ for $\nu=0.5$, $0.3$, and $0$, respectively.
In this case, the differences between the different curves are very small, but finite, as stressed by the inset.
}
\label{fig_2_FT_par}
\end{figure}

\begin{figure}
\centerline{\includegraphics[width=\columnwidth]{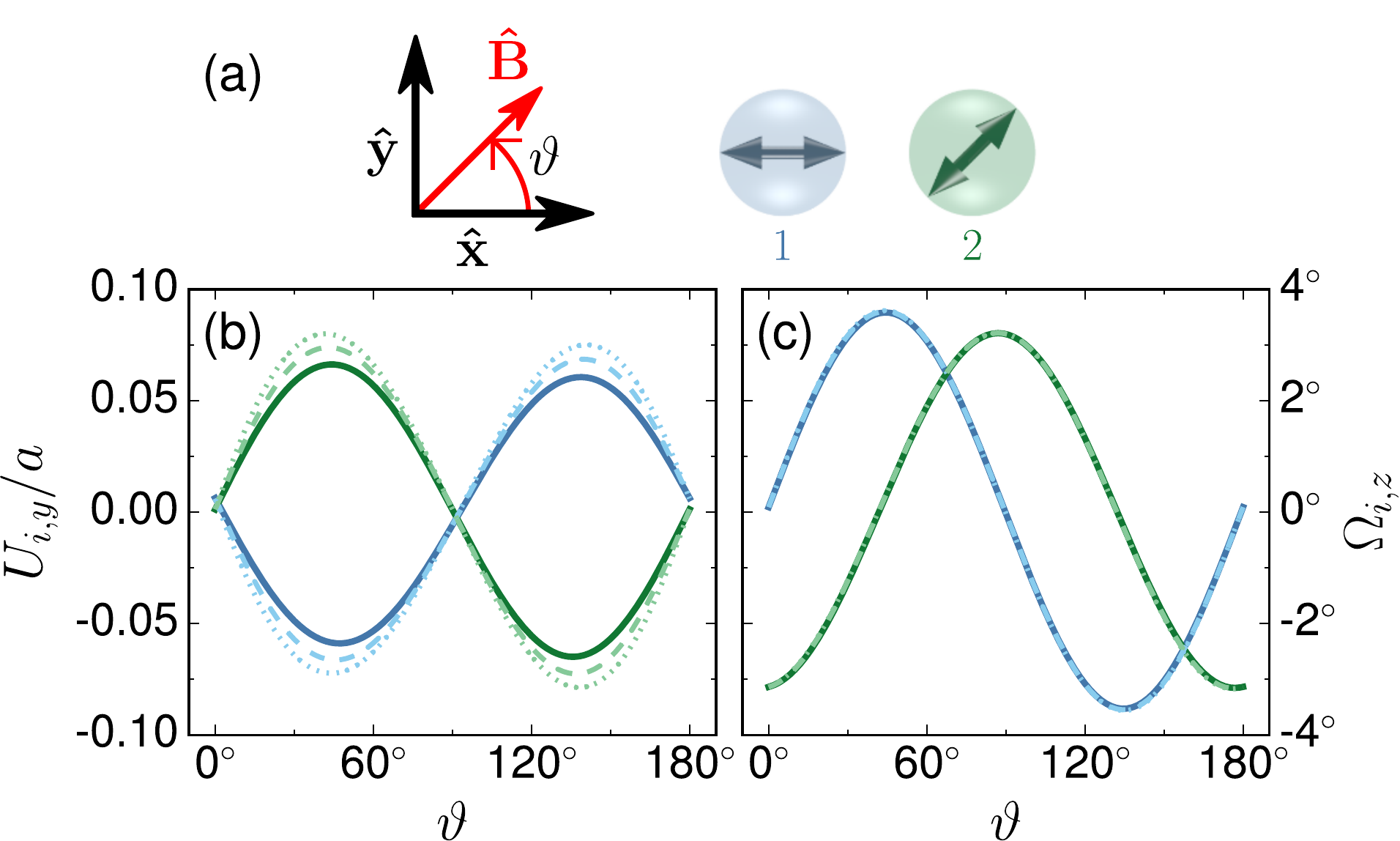}}
\caption{The same as in Fig.~\ref{fig_2_FT_par}, now with the anisotropy axis of particle 2 initially tilted for $\mathbf{B}=\mathbf{0}$ within the $xy$-plane by $45^\circ$ with respect to the $\mathbf{\hat{x}}$-axis.
}
\label{fig_2_FT}
\end{figure}

\begin{figure}
\centerline{\includegraphics[width=\columnwidth]{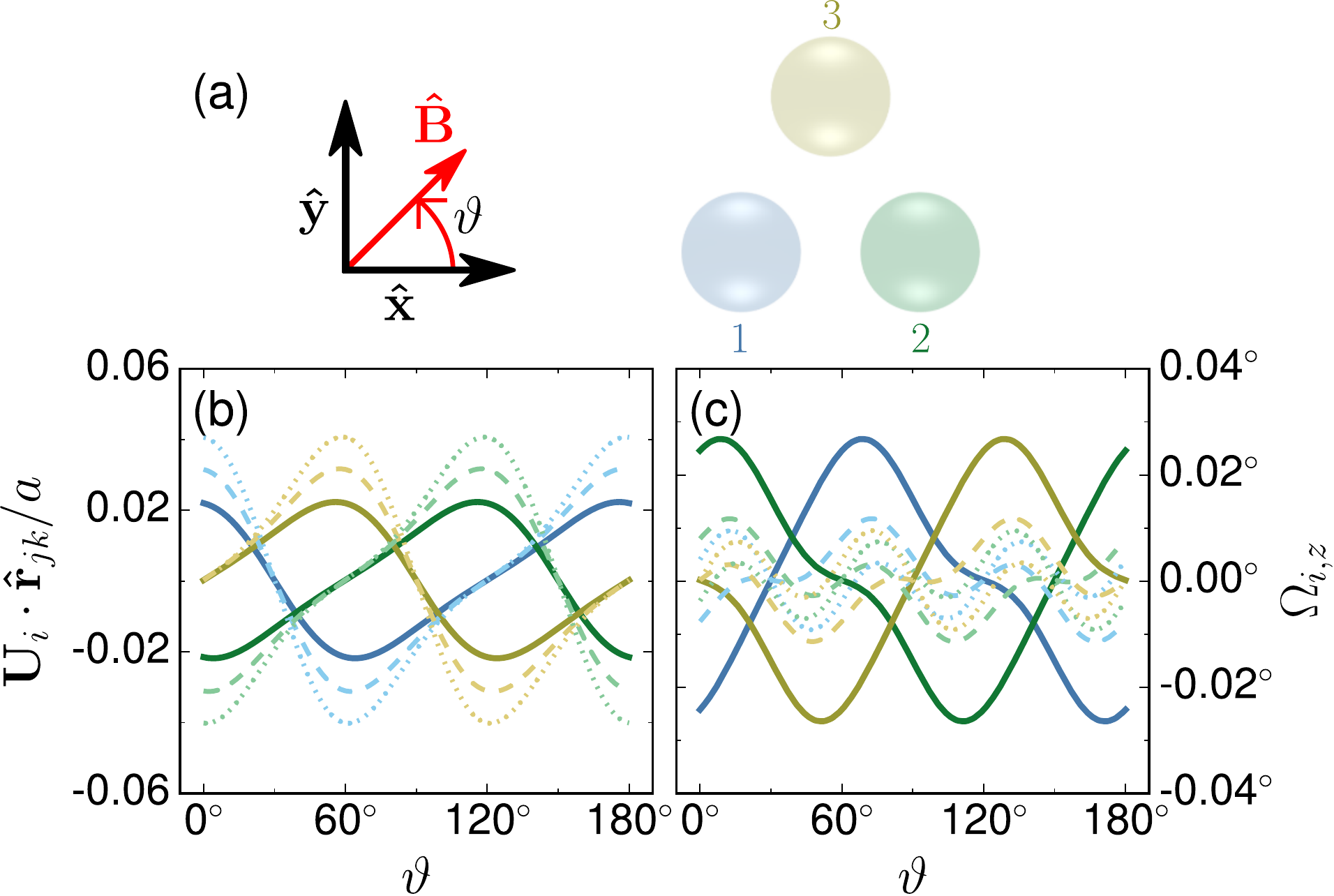}}
\caption{The same as in Fig.~\ref{fig_2_FT_par}, now for a three-particle system forming an isoscele triangle.
Here, the particles do not feature an inherent axis of anisotropy $\mathbf{\hat{n}}_i$.
Thus, only magnetic forces are induced (all magnetic torques vanish). $(i,j,k)\in\{(1,2,3),(2,3,1),(3,1,2) \}$.
}
\label{fig_3_F}
\end{figure}

\begin{figure}
\centerline{\includegraphics[width=\columnwidth]{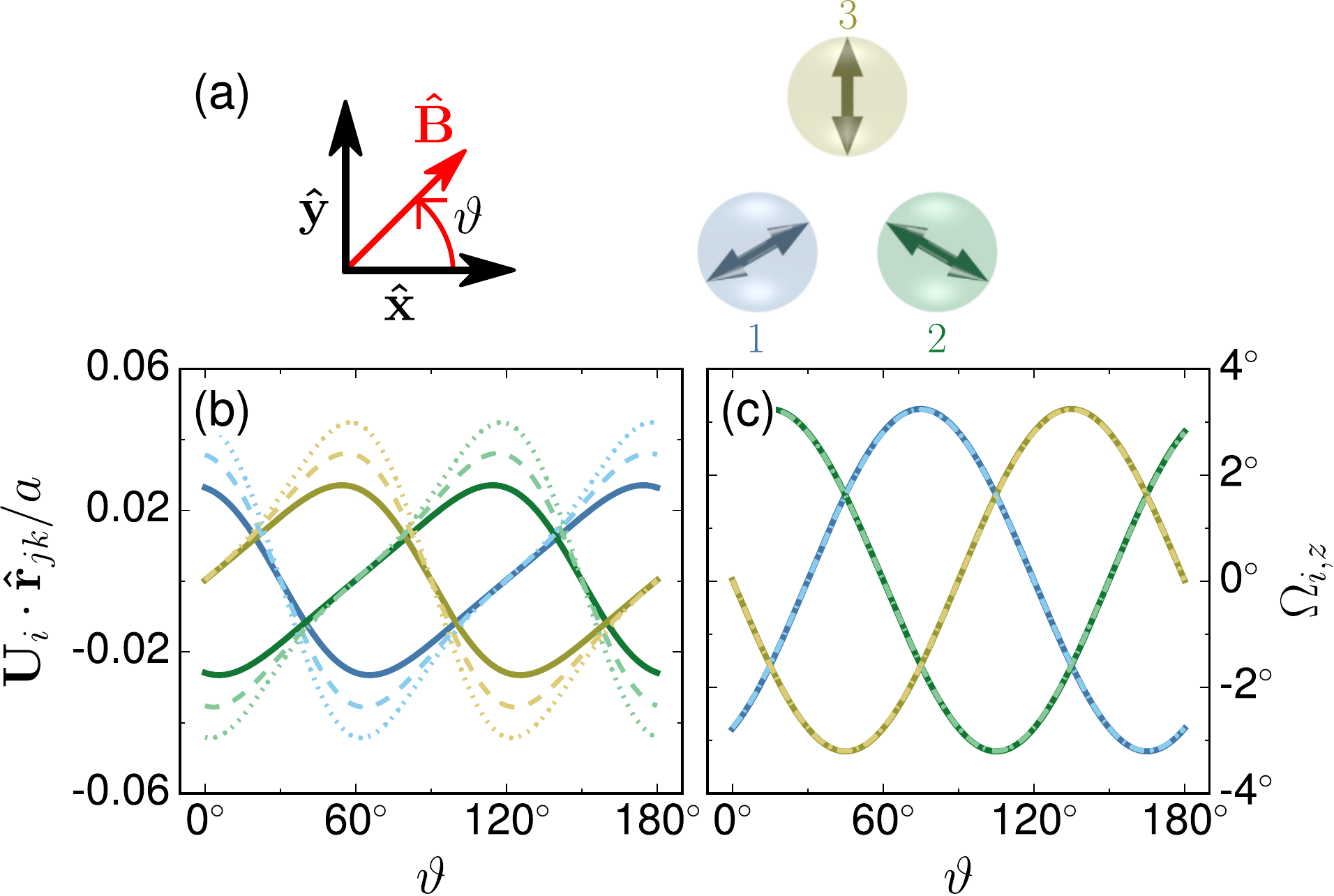}}
\caption{
The same as in Fig.~\ref{fig_3_F}, but now again with inherent axes of magnetic anisotropy $\mathbf{\hat{n}}_i$ included, oriented for $\mathbf{B}=\mathbf{0}$ as indicated by the double-headed arrows in panel (a).
Thus, both magnetic forces and torques are induced.
}
\label{fig_3_FT}
\end{figure}

In Figs.~\ref{fig_2_FT_par}--\ref{fig_3_FT}, the $xy$-plane contains the particle centers. 
The external magnetic field is rotated counterclockwise within the plane (around the $\mathbf{\hat{z}}$-axis).
Particle displacements and rotations due to the induced forces and torques are shown for $\nu=0.5$, $0.3$, and $0$ (solid, dashed, and dotted lines, respectively).
It can be seen in all examples that the translational displacements get significantly larger for decreasing $\nu$, i.e., increasing compressibility of the elastic environment.
In Figs.~\ref{fig_2_FT_par} and \ref{fig_2_FT}, both forces and torques are induced.
The differences in the rotations depicted for the different values of $\nu$, both in Figs.~\ref{fig_2_FT_par}~(c) and in \ref{fig_2_FT}~(c), are very small and here hardly visible.
This is because the Poisson ratio does not enter the $\mathbf{\underline{M}}_{ij}^\text{rt}$- and $\mathbf{\underline{M}}_{ij}^\text{rr}$-matrices below order $r_{ij}^{-5}$ and $r_{ij}^{-6}$, see Eqs.~(\ref{M_rt_5}) and (\ref{M_ij_rr_6}), respectively.

In Fig.~\ref{fig_3_F}, the particles do not feature an axis of magnetic anisotropy.
Therefore, only forces are induced by the magnetic field.
Thus the rotations in Fig.~\ref{fig_3_F}~(c) stem solely from the $\mathbf{\underline{M}}_{ij}^\text{rt}$-couplings, see Eqs.~(\ref{M_rt_0})--(\ref{M_rt_5}).
The pronounced $\nu$-dependence partly originates from the magnetic forces, which were calculated as described above.
Since for the lower considered values of $\nu$ the displacements are larger, also the changes of the magnetic forces can become larger in magnitude according to the increased changes in the mutual particle distances.
This, in turn, affects the magnitudes of the induced rotations via the $\mathbf{\underline{M}}_{ij}^\text{rt}$-matrices.
Additionally, the higher-order contributions from the $\mathbf{\underline{M}}_{ij}^\text{rt(5)}$-matrices, see Eq.~(\ref{M_rt_5}), which include three-particle interactions and which depend on $\nu$, are now more exposed.
The particular shapes of the rotation curves for $\nu=0.5$ are in fact to a larger extent affected by the $\mathbf{\underline{M}}_{ij}^\text{rt(5)}$-matrices.
For smaller $\nu$, the lower-order $\mathbf{\underline{M}}_{i\neq j}^\text{rt}$-matrices, see Eq.~(\ref{M_rt_2}), here lead to more regular rotation curves.
For comparison, the same configuration as in Fig.~\ref{fig_3_F} is repeated in Fig.~\ref{fig_3_FT}, there with permanent axes of magnetic anisotropy of the particles included.
Thus, magnetic torques are additionally induced.
The rotations in Fig.~\ref{fig_3_FT}~(c) are then again dominated by the magnetic torques via the $\mathbf{\underline{M}}_{i=j}^\text{rr}$-matrices, see Eq.~(\ref{M_i=j_rr}). 
%
%
%

Summarizing, we observe that compressibility can play a significant role for the coupling of the translations and rotations of rigid particles mediated by an embedding elastic environment.
This has at least two sources.
First, a compressible environment can allow for larger displacements of already a single particle on its own.
The resulting larger displacements can lead to stronger interactions between the particles.
Second, the coupling via the environment itself is affected directly, see the $\nu$-dependence of the expressions for the coupling matrices in Secs.~\ref{A_displaceability} and \ref{B_rotateability}.
We note that the Poisson ratio enters the rotation--translation and rotation--rotation couplings only at fifth and sixth order, respectively.
Therefore, in many cases the corresponding effects associated with compressibility of the elastic environment will be masked if torques are imposed directly on each embedded particle.

\subsection{Higher-order effects}

Already during our derivation of the additional fifth- ($r_{ij}^{-5}$) and sixth- ($r_{ij}^{-6}$) order contributions to the coupling matrices in Secs.~\ref{A_displaceability} and \ref{B_rotateability}, we have seen that qualitatively new features arise from these higher orders. Particularly, the rotation--translation, translation--rotation, and rotation--rotation couplings now become explicitly dependent on the compressibility of the embedding elastic environment. These effects are absent up to (including) the fourth order ($r_{ij}^{-4}$) considered before \cite{puljiz2017forces}. 

\begin{figure}
\centerline{\includegraphics[width=\columnwidth]{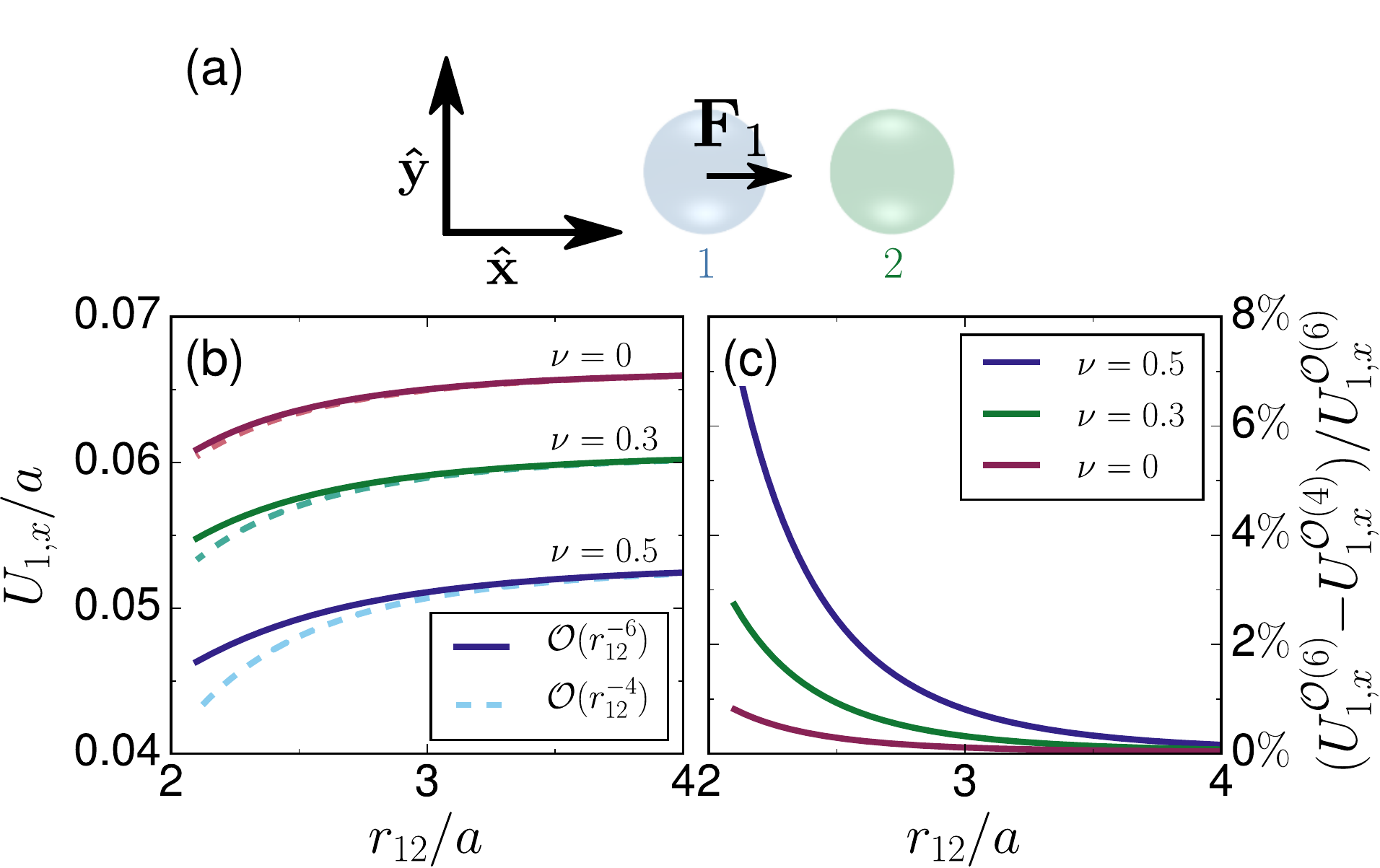}}
\caption{Two-particle system in the $xy$-plane, in which sphere 1 is pushed by a constant force $\mathbf{F}_1$ towards another rigid but ``passive" sphere.
The latter passively contributes to the net displacement of sphere 1 via the $\mathbf{\underline{M}}_{ij}^{\text{tt}(4,6)}$-matrices, see Eq.~(\ref{M_ij_tt_46}), due to its rigidity ``reflecting back" part of the displacement field generated by sphere 1.
(a) Sketch of the system configuration.
(b) The $x$-component of the displacement $\mathbf{U}_1$ of particle 1 as a function of the center-to-center interparticle distance $r_{12}$.
Dashed lines represent the contributions up to (including) order $r_{12}^{-4}$ for each considered value of $\nu$, whereas the solid lines indicate the solution up to (including) order $r_{12}^{-6}$.
(c) The relative deviation of the sixth-order $[\mathcal{O}^{(6)}]$ from the fourth-order $[\mathcal{O}^{(4)}]$ solution becomes pronounced for the considered values $\nu>0$ and for $r_{12}\lesssim 3a$.
}
\label{fig_force}
\end{figure}

\begin{figure}
\centerline{\includegraphics[width=\columnwidth]{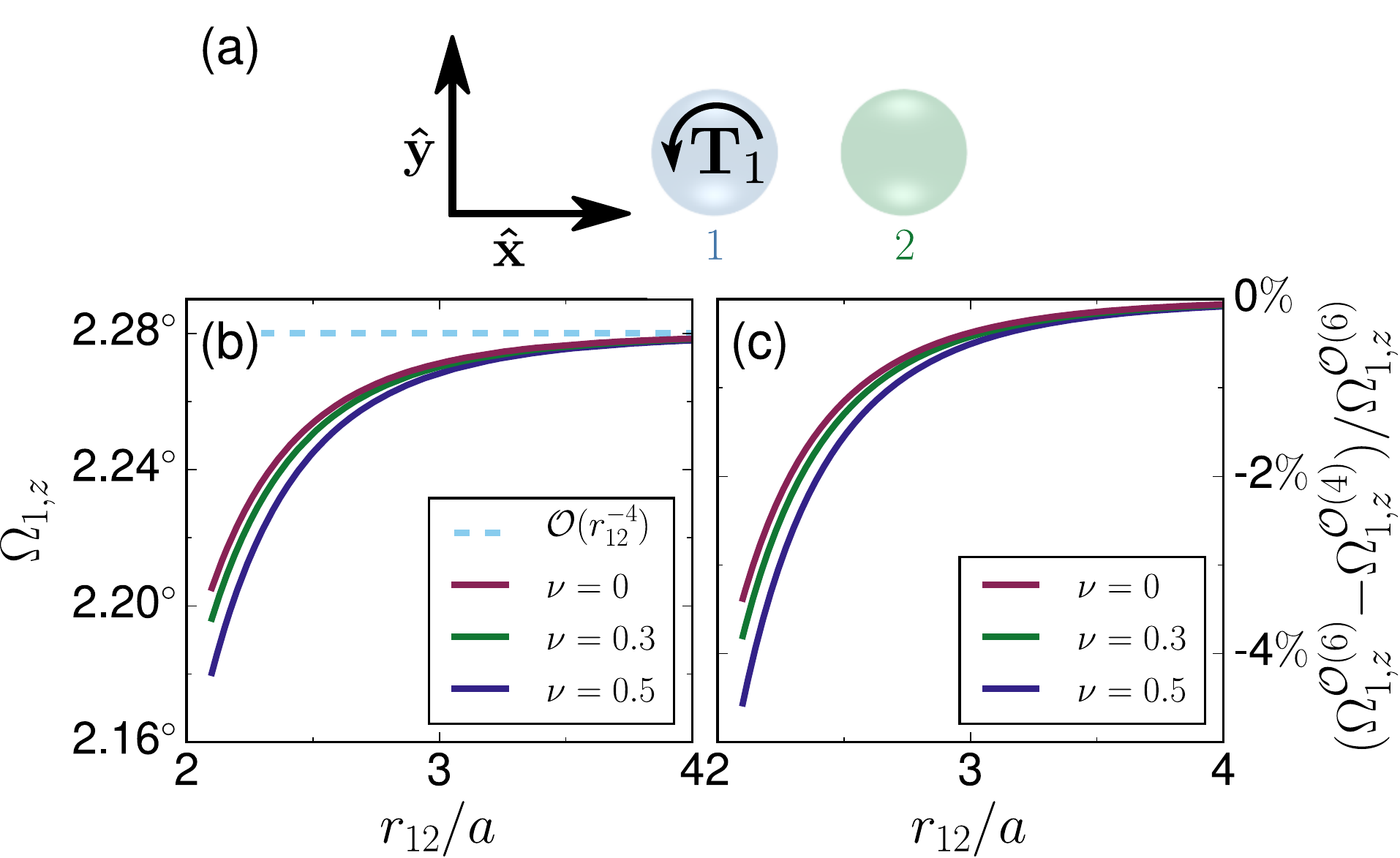}}
\caption{The same as in Fig.~\ref{fig_force}, now with an external torque $\mathbf{T}_1$ instead of a force acting on particle 1.
(b) Here, the fourth-order solution up to (including) order $r_{12}^{-4}$ of the $z$-component of the rotation $\boldsymbol{\Omega}_1$ of particle 1 as a function of the center-to-center interparticle distance $r_{12}$ equals the constant rotation of an isolated sphere (dashed line).
The solutions up to (including) order $r_{12}^{-6}$ (solid lines) show the effect of the presence of the passive sphere 2 on the net rotation of particle 1 for different values of the Poisson ratio $\nu$.
(c) Relative deviations of the sixth-order $[\mathcal{O}^{(6)}]$ from the fourth-order $[\mathcal{O}^{(4)}]$ solutions for $\Omega_{1,z}$ again become pronounced for $r_{12}\lesssim3a$.
}
\label{fig_torque}
\end{figure}

To also obtain basic quantitative numbers on the typical role of the additional contributions of orders $r_{ij}^{-5}$ and $r_{ij}^{-6}$ derived in Secs.~\ref{A_displaceability} and \ref{B_rotateability}, we consider two minimal example situations. First, one rigid sphere is subjected to a constant force $\mathbf{F}_1$ that pushes it towards another rigid, nearby, ``passive'' sphere, see Fig.~\ref{fig_force}. 
Second, one rigid sphere is rotated by a constant torque $\mathbf{T}_1$ in the presence of another rigid, nearby, ``passive'' sphere, see Fig.~\ref{fig_torque}. In practice, such a situation could be realized, for instance, by a configuration of two nearby rigid spherical particles of equal size. However, the first one of them is magnetic, the second one is not. Exposing the magnetic sphere to a magnetic field gradient, it will be subject to a net force \cite{jackson1962classical}, while already a homogeneous field misaligned with an axis of internal magnetic anisotropy will impose a torque, see Eq.~(\ref{stoner-wohlfarth}). Here, we set $|\mathbf{F}_1|=\mu a^2$ and $|\mathbf{T}_1|=\mu a^3$. We reduce the distance $r_{12}$ between the centers of the two spheres and determine the deviation of the solution up to (including) order $r_{12}^{-6}$ derived in Secs.~\ref{A_displaceability} and \ref{B_rotateability} from the previously considered \cite{puljiz2017forces} solution up to (including) order $r_{12}^{-4}$. 

In both situations, the presence of the second, ``passive'' sphere affects the translation or rotation of the first one because its rigidity opposes to deformations in the environment generated by the first sphere. Accordingly, part of the displacement field induced by the first sphere is ``reflected back'' towards the latter. Concerning the results depicted in Fig.~\ref{fig_force} for the case of an imposed force, this effect is included by the $\mathbf{\underline{M}}_{ij}^{\text{tt}(4,6)}$-matrices, see Eq.~(\ref{M_ij_tt_46}).
There are already fourth-order contributions ($r_{12}^{-4}$) of this kind, with further terms arising to sixth order ($r_{12}^{-6}$).
Obviously, when decreasing the distances $r_{12}$ towards $2a$, the sixth-order contributions become more pronounced.
Moreover, for the considered values of the Poisson ratio $\nu=0$, $0.3$, and $0.5$, the strongest relative deviation between the sixth- and fourth-order solutions is found for $\nu=0.5$, see Fig.~\ref{fig_force} (c).

Turning to the situation of an imposed torque on the first particle, see Fig.~\ref{fig_torque}, no reflections of the kind described above occur up to (including) the fourth order, to which the rotation $\boldsymbol{\Omega}_1$ of sphere 1 is independent of $r_{12}$, see Eq.~(\ref{M_i=j_rr}).
[Since there is no magnetic torque acting on sphere 2, the $\mathbf{\underline{M}}_{i\neq j}^\text{rr}$-matrices given by Eq.~(\ref{M_ineqj_rr}) do not contribute to $\boldsymbol{\Omega}_1$.]
Nevertheless, the presence of the rigid second sphere is felt by the first sphere to sixth order ($r_{12}^{-6}$), see Eq.~(\ref{M_ij_rr_6}).
Moreover, a slightly different behavior is found for the different considered values of $\nu$, see Fig.~\ref{fig_torque} (b) and (c).
The observed attenuation of the $z$-component of $\boldsymbol{\Omega}_1$ is most pronounced for $\nu=0.5$, and it grows in magnitude when decreasing the interparticle center-to-center distance $r_{12}$ towards $2a$.

Altogether, we observe in our examples that the higher-order contributions derived above become increasingly relevant when the Poisson ratio is growing from $\nu=0$ towards the limit of incompressibility of $\nu=0.5$. Moreover, the additional contributions notably grow when decreasing the center-to-center interparticle distance below approximately $3a$. For larger interparticle distances, fourth-order solutions for the particle displacements and rotations appear to be already of a reasonable quantitative precision \cite{puljiz2016forces,puljiz2018reversible}. At some point, when the particles come too close to each other, depending on the particular material and system investigated, induced heterogeneities around the particles \cite{huang2016buckling} or nonlinear elastic effects \cite{puljiz2018reversible} may become important.


\section{Conclusions}
\label{conclusion}

In summary, we have presented a modified route of calculating both the classical solution of the displacement field and the incompressible Stokes flow around a translated and rotated rigid sphere of no-slip surface conditions in a surrounding possibly compressible linearly elastic solid or incompressible low-Reynolds-number fluid, respectively.
(The latter result formally follows from the former by considering the limit of an incompressible surrounding medium.)
For this purpose, the displacement field in an elastic environment has been expanded around the spherical particle. 
Then an explicit ansatz has been inserted for the surface force density as also presented, e.g., in Ref.~\onlinecite{dhont1996introduction}.
Afterwards, we showed by explicit calculation that the ansatz renders the expansion finite.

Next, the two Fax\'en laws beyond the stresslet describing the reaction of the sphere to an imposed deformation or flow field in its environment were derived, both for linear elasticity theory (within a possibly compressible elastic solid) and for low-Reynolds-number hydrodynamics (within an incompressible fluid), respectively.

Based on these results, we calculated the displaceability and rotateability matrices up to the sixth order in inverse particle separation distance, characterizing the particle interactions mediated by the surrounding compressible elastic environment.
Differences when compared to the incompressible situation were demonstrated via basic example situations, together with a brief illustration of the effects of the additional higher-order contributions derived in the present work.

Our results will be important in the future, for instance, to quantify to higher-order precision the properties and behavior of complex elastic composite materials and their response to external stimuli \cite{filipcsei2007magnetic,menzel2015tuned,odenbach2016microstructure,puljiz2018reversible}.
Since, frequently, in soft complex elastic composite or multi-component materials a pronounced nonlinear stress-strain behavior is strongly dominated by interactions resulting from the additional components \cite{menzel2009nonlinear,cremer2016superelastic} and not necessarily by the nonlinear elasticity of the purely elastic component, our results derived in the framework of linear elasticity theory may to some extent even serve to characterize more in detail the nonlinear stress-strain behavior of such systems.

\begin{acknowledgments}
The authors thank the Deutsche Forschungsgemeinschaft (DFG) for support of this work through the priority program SPP 1681, Grant No.~ME 3751/3. 
\end{acknowledgments}

\appendix

\begin{widetext}
\section*{Appendix A}\label{Appendix_M}
\renewcommand{\theequation}{A.\arabic{equation}}
\setcounter{equation}{0}

\begin{table*}
	    \begin{eqnarray*}
	     	\int_{\partial V} \mathrm{d}S(\mathbf{\hat{k}})~\Psi_{kl} &={}&  \frac{4\pi}{3}\delta_{kl}\\
	     	\int_{\partial V} \mathrm{d}S(\mathbf{\hat{k}})~\Psi_{kl}
	     	\hat{k}_m\hat{k}_n
	     	&={} &\frac{2\pi }{15}\left[4\,\delta_{kl}\delta_{mn} - (\delta_{km}\delta_{ln} + \delta_{kn}\delta_{lm})\right]
	     		\\
	     	\int_{\partial V} \mathrm{d}S(\mathbf{\hat{k}})~
	     	\hat{k}_i\hat{k}_j
	     	\Psi_{kl}
	     	\hat{k}_m\hat{k}_n
	     	&={} &\frac{2\pi}{105}\Big\{
	     	6\,\delta_{kl}(\delta_{ij}\delta_{mn}+\delta_{im}\delta_{jn}+\delta_{in}\delta_{jm})\notag\\
	     	&{}&-\Big[
	     		\delta_{ij}(\delta_{km}\delta_{ln}+\delta_{kn}\delta_{lm}) + \delta_{mn}(\delta_{ik}\delta_{jl}+\delta_{il}\delta_{jk}) + \delta_{ik}(\delta_{jm}\delta_{ln}+\delta_{jn}\delta_{lm})\\
	     		&{}&
	     		+\delta_{jk}(\delta_{im}\delta_{ln}+\delta_{in}\delta_{lm})
	     		+\delta_{km}(\delta_{il}\delta_{jn}+\delta_{in}\delta_{jl})
	     		+\delta_{kn}(\delta_{il}\delta_{jm}+\delta_{im}\delta_{jl})
	     	\Big]
	     	\Big\}
	  	     	\\
	       	\int_{\partial V} \mathrm{d}S(\mathbf{\hat{k}})~
	     	 	\hat{k}_i\hat{k}_j
	     	   	\Psi_{kl}
	     	   	\hat{k}_m\hat{k}_n\hat{k}_o\hat{k}_p
	     	   	&={} &\frac{2\pi}{945}\Big\{
	     	   	8\,\delta_{kl}(\delta_{ij}\delta_{mn}\delta_{op}+\text{all permutations}) 
	     	   	\notag\\
	     	   		     	   	&{}&
	     	   	- \Big[
	     	   	\delta_{ij}\delta_{km}\delta_{ln}\delta_{op}+
	     	   	\text{all permutations except for those containing }\delta_{kl}
	     	   	\Big]
	     	   	\Big\}
	     \end{eqnarray*}
	\caption{
		Here, $\Psi_{kl}:=(\hat{\vartheta}_k\hat{\vartheta}_l+\hat{\varphi}_k\hat{\varphi}_l)/2$.
		For $k=l$, $\Psi_{kk}=1$ and the results from Tab.~\ref{table_int} are recovered.
		In the first term on the right-hand side of the last equation, all permutations of $\delta_{ij}\delta_{mn}\delta_{op}$ are added (i.e., $\delta_{ij}\delta_{mo}\delta_{np}$, $\delta_{ij}\delta_{mp}\delta_{no}$, and so on), which makes in total 15 terms. 
		The second term on this right-hand side is a superposition of all permutations of $\delta_{ij}\delta_{km}\delta_{ln}\delta_{op}$, except for those containing $\delta_{kl}$, thus in total 90 terms (altogether, the number of terms on the right-hand side is therefore $15+90=105$).
	}
	\label{table_int2}
\end{table*}

The integral on the right-hand side of Eq.~(\ref{boundary_mjkl}),
\begin{equation}
	\int_{\partial V}\mathrm{d}S'\int_{\partial V}\mathrm{d}S~r_kr_l\, G_{ij}(\mathbf{r}-\mathbf{r}')f_j(\mathbf{r}'),
\end{equation}
can be solved by Fourier forth and back transform.
The Fourier transform of the Green's function reads \cite{puljiz2017forces}
\begin{equation}
	\mathbf{\tilde{\underline{G}}}(\mathbf{k}) ={} \frac{1}{\mu k^2}\left[
		\mathbf{\hat{\underline{I}}} - \frac{1}{2(1-\nu)}\mathbf{\hat{k}}\mathbf{\hat{k}}
	\right],
\end{equation}
where we introduced the wave vector $\mathbf{k}$, $k=|\mathbf{k}|$, and $\mathbf{\hat{k}}=\mathbf{k}/k$.
First, only the $\mathrm{d}S$-integral is considered,
\begin{equation}
	(2\pi)^3\int_{\partial V}\mathrm{d}S~r_kr_l\, G_{ij}(\mathbf{r}-\mathbf{r}') ={} 
	\int_{\partial V}\mathrm{d}S\int_{\mathbb{R}}\mathrm{d}^3k\frac{1}{\mu k^2}
	\left[\delta_{ij}-\frac{1}{2(1-\nu)}\hat{k}_i\hat{k}_j\right]r_k r_l e^{i\mathbf{k}\cdot(\mathbf{r}-\mathbf{r}')}.\qquad
\end{equation}
Using the identity
\begin{equation}
	r_{k}r_l e^{i\mathbf{k}\cdot\mathbf{r}} ={} -\nabla_k^\mathbf{k}\nabla_l^\mathbf{k}e^{i\mathbf{k}\cdot\mathbf{r}},
\end{equation}
where
\begin{equation}
	\nabla_l^\mathbf{k} ={} \hat{k}_l\frac{\partial}{\partial k} +
	\hat{\vartheta}_l \frac{1}{k}\frac{\partial}{\partial \vartheta}
	+\hat{\varphi}_l \frac{1}{k \sin(\vartheta)}\frac{\partial}{\partial \varphi},
\end{equation}
and
\begin{equation}
	\int_{\partial V}\mathrm{d}S~e^{i\mathbf{k}\cdot\mathbf{r}} ={} 4\pi a^2 \frac{\sin(ka)}{ka},
\end{equation}
we obtain
\begin{eqnarray}\label{app_a_dS}
	\lefteqn{\int_{\partial V}\mathrm{d}S~r_kr_l\, G_{ij}(\mathbf{r}-\mathbf{r}')}\notag \\&={}&
	-\frac{a^2}{2\pi^2 \mu}\int_{\partial V}\mathrm{d}S(\mathbf{\hat{k}})
	\left[\delta_{ij}-\frac{1}{2(1-\nu)}\hat{k}_i\hat{k}_j\right]
	\int_{0}^\infty\mathrm{d}k~ e^{-i\mathbf{k}\cdot\mathbf{r}'}\left\{
	\hat{k}_k\hat{k}_l \frac{\mathrm{d}^2}{\mathrm{d}k^2}
	+(\hat{\vartheta}_k\hat{\vartheta}_l + \hat{\varphi}_k\hat{\varphi}_l) \frac{1}{k}\frac{\mathrm{d}}{\mathrm{d}k}
	\right\} \frac{\sin(ka)}{ka}.
\end{eqnarray}
After solving the $\mathrm{d}k$-integral, only the part even in $\mathbf{\hat{k}}$ can survive the subsequent $\mathrm{d}S(\mathbf{\hat{k}})$-integral.
Thus, the limits of integration may here be rewritten as $\int_{0}^{\infty}\mathrm{d}k=\frac{1}{2}\int_{\mathbb{R}}\mathrm{d}k$.
With
\begin{eqnarray}
	\frac{1}{2}\int_{\mathbb{R}}\mathrm{d}k\frac{1}{k}e^{-i k \mathbf{\hat{k}}\cdot\mathbf{r}'} \frac{\mathrm{d}}{\mathrm{d} k} \frac{\sin(ka)}{ka}&={} & \frac{\pi}{4 a}(\mathbf{\hat{k}}\cdot\mathbf{r}')^2-\frac{\pi a}{4},	\label{app_int_dk}\\
	\frac{1}{2}\int_{\mathbb{R}}\mathrm{d}ke^{-i k \mathbf{\hat{k}}\cdot\mathbf{r}'} \frac{\mathrm{d}^2}{\mathrm{d} k^2} \frac{\sin(ka)}{ka}&={} & -\frac{\pi}{2 a}(\mathbf{\hat{k}}\cdot\mathbf{r}')^2, \label{app_int_dk2}
\end{eqnarray}
for $-1<\mathbf{\hat{k}}\cdot\mathbf{r}'/a<1$, which can be obtained by Mathematica \cite{ram2010} as well as by partial integration and further proceeding as outlined in the appendices of Refs.~\onlinecite{dhont1996introduction} and \onlinecite{puljiz2017forces}, respectively, Eq.~(\ref{app_a_dS}) becomes
\begin{eqnarray}
	\lefteqn{\int_{\partial V}\mathrm{d}S~r_kr_l\, G_{ij}(\mathbf{r}-\mathbf{r}')} \notag\\
	&={} &
		\frac{a}{4\pi \mu}\int_{\partial V}\mathrm{d}S(\mathbf{\hat{k}})
		\left[\delta_{ij}-\frac{1}{2(1-\nu)}\hat{k}_i\hat{k}_j\right]
		\left\{
			\hat{k}_k\hat{k}_l\hat{k}_m\hat{k}_n r_m'r_n'
			+\frac{1}{2}(\hat{\vartheta}_k\hat{\vartheta}_l+\hat{\varphi}_k\hat{\varphi}_l)a^2
			-\frac{1}{2}(\hat{\vartheta}_k\hat{\vartheta}_l+\hat{\varphi}_k\hat{\varphi}_l) \hat{k}_m\hat{k}_n r_m'r_n'
		\right\}.\qquad\quad
\end{eqnarray}
Evaluation with the help of Tabs.~\ref{table_int} and \ref{table_int2} directly leads to Eq.~(\ref{int_m_jkl}).

\section*{Appendix B}\label{Appendix_N}
\renewcommand{\theequation}{A.\arabic{equation}}
\setcounter{equation}{0}

The same procedure as in Appendix A is applied to evaluate the integral
\begin{equation}
	\int_{\partial V}\mathrm{d}S'\int_{\partial V}\mathrm{d}S~r_kr_lr_m\, G_{ij}(\mathbf{r}-\mathbf{r}')f_j(\mathbf{r}'),
\end{equation}
now using the identity
\begin{equation}
	r_{k}r_lr_m e^{i\mathbf{k}\cdot\mathbf{r}} ={} i\nabla_k^\mathbf{k}\nabla_l^\mathbf{k}\nabla_m^\mathbf{k}e^{i\mathbf{k}\cdot\mathbf{r}}.
\end{equation}
This leads to
%
\begin{eqnarray}
	\lefteqn{\int_{\partial V}\mathrm{d}S~r_kr_lr_m\, G_{ij}(\mathbf{r}-\mathbf{r}')}\notag\\ &={} & 
	i\frac{a^2}{2\pi^2 \mu}\int_{\partial V}\mathrm{d}S(\mathbf{\hat{k}})
	\left[\delta_{ij}-\frac{1}{2(1-\nu)}\hat{k}_i\hat{k}_j\right]
	\int_{0}^\infty\mathrm{d}k~ e^{-i\mathbf{k}\cdot\mathbf{r}'}\bigg\{
	\hat{k}_k\hat{k}_l\hat{k}_m \frac{\mathrm{d}^3}{\mathrm{d}k^3}\notag\\
	&{}&
	+(
		\hat{k}_k\hat{\vartheta}_l\hat{\vartheta}_m
		+ \hat{k}_k\hat{\varphi}_l\hat{\varphi}_m
		+ \hat{\vartheta}_k\hat{\vartheta}_l\hat{k}_m
		+ \hat{\vartheta}_k\hat{k}_l\hat{\vartheta}_m
		+ \hat{\varphi}_k\hat{\varphi}_l\hat{k}_m
		+ \hat{\varphi}_k\hat{k}_l\hat{\varphi}_m
	) \left(\frac{1}{k}\frac{\mathrm{d}^2}{\mathrm{d}k^2}
	-\frac{1}{k^2}\frac{\mathrm{d}}{\mathrm{d}k}\right)
	\bigg\} \frac{\sin(ka)}{ka}.
\end{eqnarray}
As in Appendix A, we may equivalently use $\int_{0}^{\infty}\mathrm{d}k=\frac{1}{2}\int_{\mathbb{R}}\mathrm{d}k$ here.
Integrating by parts, we find
\begin{eqnarray}
	\frac{1}{2}\int_{\mathbb{R}}\mathrm{d}k~e^{-i k \mathbf{\hat{k}}\cdot\mathbf{r}'}\frac{\mathrm{d}^3}{\mathrm{d}k^3}\frac{\sin(ka)}{ka} &={}&  \frac{1}{2}i(\mathbf{\hat{k}}\cdot\mathbf{r}')\int_\mathbb{R}\mathrm{d}k~e^{-ik\mathbf{\hat{k}}\cdot\mathbf{r}'}\frac{\mathrm{d}^2}{\mathrm{d}k^2}\frac{\sin(ka)}{ka}, \\
	\frac{1}{2}\int_{\mathbb{R}}\mathrm{d}k~e^{-i k \mathbf{\hat{k}}\cdot\mathbf{r}'}
	\left(\frac{1}{k}\frac{\mathrm{d}^2}{\mathrm{d}k^2}-\frac{1}{k^2}\frac{\mathrm{d}}{\mathrm{d}k}\right)\frac{\sin(ka)}{ka}&={}&\frac{1}{2}
	i(\mathbf{\hat{k}}\cdot\mathbf{r}')\int_{\mathbb{R}}\mathrm{d}k~e^{-ik\mathbf{\hat{k}}\cdot\mathbf{r}'}\frac{1}{k}\frac{\mathrm{d}}{\mathrm{d}k}\frac{\sin(ka)}{ka}.
\end{eqnarray}
The remaining integrals are the same as in Eqs.~(\ref{app_int_dk2}) and (\ref{app_int_dk}), respectively.
Altogether, we obtain
\begin{eqnarray}
	\lefteqn{\int_{\partial V}\mathrm{d}S~r_kr_lr_m\, G_{ij}(\mathbf{r}-\mathbf{r}')} \notag\\
	&={} &
		\frac{a}{4\pi \mu}\int_{\partial V}\mathrm{d}S(\mathbf{\hat{k}})
		\left[\delta_{ij}-\frac{1}{2(1-\nu)}\hat{k}_i\hat{k}_j\right]
		\bigg\{
			\hat{k}_k\hat{k}_l\hat{k}_m\hat{k}_n \hat{k}_o\hat{k}_p
			r_n'r_o'r_p'\notag\\
		&{}&
			+\frac{1}{2}(\hat{\vartheta}_k\hat{\vartheta}_l+\hat{\varphi}_k\hat{\varphi}_l) \hat{k}_m \hat{k}_n r_n' a^2
			+\frac{1}{2}(\hat{\vartheta}_k\hat{\vartheta}_m+\hat{\varphi}_k\hat{\varphi}_m) \hat{k}_l \hat{k}_n r_n' a^2
			+\frac{1}{2}(\hat{\vartheta}_l\hat{\vartheta}_m+\hat{\varphi}_l\hat{\varphi}_m) \hat{k}_k \hat{k}_n r_n' a^2
			\notag\\
			&{}&
			-\frac{1}{2}(\hat{\vartheta}_k\hat{\vartheta}_l+\hat{\varphi}_k\hat{\varphi}_l) \hat{k}_m\hat{k}_n\hat{k}_o\hat{k}_p r_n'r_o'r_p'
			-\frac{1}{2}(\hat{\vartheta}_k\hat{\vartheta}_m+\hat{\varphi}_k\hat{\varphi}_m) \hat{k}_l\hat{k}_n\hat{k}_o\hat{k}_p r_n'r_o'r_p'
			-\frac{1}{2}(\hat{\vartheta}_l\hat{\vartheta}_m+\hat{\varphi}_l\hat{\varphi}_m) \hat{k}_k\hat{k}_n\hat{k}_o\hat{k}_p r_n'r_o'r_p'
		\bigg\},\quad\qquad
\end{eqnarray}
finally resulting in Eq.~(\ref{N_eq}) after evaluation of the integrals using Tabs.~\ref{table_int} and \ref{table_int2}.
\end{widetext}


%

\end{document}